\documentclass[journal,10pt]{IEEEtran}
\usepackage[pdftex]{graphicx}
\usepackage[caption=false,font=footnotesize]{subfig}
\usepackage{ifpdf}
\usepackage{cite}
\usepackage{amsmath}
\usepackage{mdwmath}
\usepackage{mdwtab}
\usepackage{array}
\usepackage{stfloats}
\usepackage{url}
\usepackage{color,xcolor}
\usepackage{siunitx}
\usepackage{footnote}
\usepackage{booktabs} 
\usepackage{threeparttable}
\usepackage{multirow}
\usepackage{mathrsfs}
\usepackage{amsthm,amsmath,amssymb}

\begin{document}
%
\title{Boosting Personalised Musculoskeletal Modelling with Physics-informed Knowledge Transfer}
%
%

\author{Jie Zhang,~\IEEEmembership{Member, IEEE},
        Yihui Zhao,
        Tianzhe Bao,
        Zhenhong Li,~\IEEEmembership{Member, IEEE},
        Kun Qian,
        Alejandro F. Frangi,~\IEEEmembership{Fellow, IEEE},
        Sheng Quan Xie,~\IEEEmembership{Senior Member, IEEE},
        Zhi-Qiang Zhang,~\IEEEmembership{Member, IEEE}
\thanks{This work was supported in part by U.K. EPSRC under Grant
EP/S019219/1 and Grant EP/V057782/1, and in part by EU Marie Curie Individual Fellowship under grant 101023097. For the purpose of open access, the author(s) has applied a Creative Commons Attribution (CC BY) licence (where permitted by UKRI, ‘Open Government Licence’ or ‘Creative Commons Attribution No-derivatives (CC BY-ND) licence’ may be stated instead) to any Author Accepted Manuscript version arising. (Corresponding authors: Zhi-Qiang Zhang and Alejandro F. Frangi)} 
\thanks{Jie Zhang is with the School of Electronic and Electrical Engineering, University of Leeds, Leeds, LS2 9JT, U.K., and also with the School of Automation and Electrical Engineering, University of Science and Technology Beijing, Beijing, 100083, China (email: eenjz@leeds.ac.uk).}
\thanks{Yihui Zhao, Zhenhong Li, Kun Qian, Sheng Quan Xie, and Zhi-Qiang Zhang are with the School of Electronic and Electrical Engineering, University of Leeds, Leeds, LS2 9JT, U.K. (e-mails: \{eenyzhao, z.h.li, fbskq, s.q.xie, z.zhang3\}@leeds.ac.uk).}
\thanks{Tianzhe Bao is with the School of Rehabilitation Sciences and Engineering, University of Health and Rehabilitation Sciences, Qingdao, 261000, China (e-mail: tianzhe.bao@uor.edu.cn).}
\thanks{Alejandro F. Frangi is with the School of Computing, University of Leeds, Leeds, LS2 9JT, U.K., also with the Alan Turing Institute, London, NW1 2DB, U.K., and also with the Department of Electrical Engineering, KU Leuven, 3000 Leuven, Belgium (e-mail: a.frangi@leeds.ac.uk).}
}

\maketitle

\begin{abstract}
Data-driven methods have become increasingly more prominent for musculoskeletal modelling due to their conceptually intuitive simple and fast implementation. However, the performance of a pre-trained data-driven model using the data from specific subject(s) may be seriously degraded when validated using the data from a new subject, hindering the utility of the personalised musculoskeletal model in clinical applications. This paper develops an active physics-informed deep transfer learning framework to enhance the dynamic tracking capability of the musculoskeletal model on the unseen data. The salient advantages of the proposed framework are twofold: 1) For the generic model, physics-based domain knowledge is embedded into the loss function of the data-driven model as soft constraints to penalise/regularise the data-driven model. 2) For the personalised model, the parameters relating to the feature extraction will be directly inherited from the generic model, and only the parameters relating to the subject-specific inference will be fine-tuned by jointly minimising the conventional data prediction loss and the modified physics-based loss. In this paper, we use the synchronous muscle forces and joint kinematics prediction from surface electromyogram (sEMG) as the exemplar to illustrate the proposed framework. Moreover, convolutional neural network (CNN) is employed as the deep neural network to implement the proposed framework, and the physics law between muscle forces and joint kinematics is utilised as the soft constraints. Results of comprehensive experiments on a self-collected dataset from eight healthy subjects indicate the effectiveness and great generalization of the proposed framework.     
\end{abstract}

%
\IEEEpeerreviewmaketitle

\section{Introduction}
\IEEEPARstart{C}{omputational} musculoskeletal modelling aims to understand the mechanism of the nervous system to learn and adapts to physiological modifications, which is critical for various clinical applications, such as planning rehabilitative treatments, prostheses and robotics control, and designing assistive devices~\cite{li2022new,bennett2022emg,weng2022adaptive}. Physics-based musculoskeletal modelling methods could interpret transformation among neural excitation, muscle dynamics, and joint kinematics and kinetics using experimental measurements from markers and sensors~\cite{jung2021intramuscular,zhao2020emg}. Physiological quantities, e.g., muscle forces and joint moment, could be successively estimated with electromyograms (EMGs), foot-ground reaction forces (GRFs), and segmental body kinematics, etc. However, such set of methods is time-consuming and slow, especially for complex models in high-dimensional spaces, hindering the large-scale implementation in real-time application scenarios~\cite{xiong2021deep}.

To tackle the slowness of physics-based musculoskeletal modelling methods, machine/deep learning-based data-driven methods are recently used to build mappings between EMGs, and muscle forces/joint kinematics~\cite{rane2019deep,de2020machine,makaram2021analysis,zhang2020robust,bao2020cnn} to reduce the time consumption in musculoskeletal model building~\cite{bao2022towards}. For instance, Tang et al.~\cite{tang2021decoding} developed a modified framework to estimate the muscle force using surface EMG (sEMG) based on an encoder-decoder network. Wimalasena et al.~\cite{wimalasena2022estimating} proposed a large-scale unsupervised deep learning method to achieve the spatial and temporal representations of the multi-muscle activation from EMG measurements. Burton et al.~\cite{burton2021machine} employed four machine/deep learning methods, including convolutional neural network (CNN), recurrent neural network (RNN), fully-connected neural network, and principal component regression, to estimate the lower extremity muscle and joint contact forces from joint kinematics, GRFs, and anthropometrics. {However, all these data-driven methods would require a large number of training data to make the model  personalised. The performance of a pre-trained data-driven model using the data from specific subject(s) may be seriously degraded when validated using the data from a new subject.}

The targeted outputs of the data-driven model are normally derived from the physics-based model using static optimisation, and it is very time-consuming to get a large number of training data. To overcome this issue, transfer learning has been explored recently to investigate how to make a pre-trained model work for a new subject with only limited training data from him/her. Dao et al.~\cite{dao2019deep} developed a deep transfer learning strategy based on long short-term memory (LSTM) to predict skeletal muscle forces from kinematic measurements during a gait cycle. Bao et al.~\cite{bao2021inter} proposed a novel two-stream CNN for supervised domain adaptation to reduce domain shift effects on wrist kinematics estimation in the inter-subject circumstance. Wang et al.~\cite{wang2018sensor} utilised EMG signals and acceleration data as the multimodal input of the deep learning model to enhance the adaptability to the effects of arm movements, and transfer learning was considered to accelerate the convergence of deep learning model and avoid over-fitting problem. Kim et al.~\cite{kim2019subject} proposed a modified subject-transfer framework, in which supportive CNN classifiers were pre-trained using EMG signals from several subjects, and then transfer learning was utilised to fine-tune these classifiers to enhance the robustness of the proposed method in terms of intra-user variability. However, state-of-the-art methods are actually ``black-box", and its narrow focus on the input-output mappings may be inherently deterministic without considering the explicit physics modelling of the underlying neuromechanical processes~\cite{sartori2016neural,karniadakis2021physics,raissi2019physics,willard2021integrating,zhao2022adaptive,zhang2020non}. 
  
To tackle the aforementioned issues, a physics-informed deep transfer learning framework for personalised musculoskeletal modelling is proposed to learn the mappings from EMGs to muscle forces and joint kinematics in this paper. The proposed framework consists of a generic network and a personalised network. Specifically, a generic network is first trained with sEMG measurements from several subjects by minimising the loss function, which jointly considers the minimisation of the conventional data prediction loss and a physics-based loss. The parameters relating to the generic information of subjects are then directly shared to the personalised network as the backbone. After that, the parameters relating to the subject-specific information in the personalised network could be achieved with limited available subject-specific data based on the modified loss function. In this manner, the derived personalised network will contain mappings between kinematic measurements, and internal forces and muscle activations with the shared parameters from the generic network, which could help the personalised network hold the generalization of the generic network achieved from subjects’ data. More importantly, with many parameters frozen, over-fitting problem could be avoided during the knowledge transfer phase. sEMG-based wrist muscle forces and joint kinematics estimation is considered as the exemplar to verify the feasibility of the proposed framework. Additionally, CNN is employed as the deep neural network to implement the proposed framework. {The main contributions can be summarized as follows:}

(1) A deep learning and physical knowledge integration-based knowledge transfer framework for personalised musculoskeletal modelling is developed. The twin neural networks architecture, i.e., the generic network and the personalised network, could significantly reduce the training data required for any new subject.

(2) A modified loss function is designed by embedding physics law, in which the physics-based domain knowledge is utilised as soft constraints to penalise/regularise the loss function to enhance the robustness and prediction performance of deep neural networks, and make the intermediate functional relationships of the deep learning-based “black-box” modelling be reflected and controlled by the underlying physical mechanisms.

(3) Comprehensive experiments on a self-collected dataset from eight healthy subjects are performed, and the results demonstrate that the proposed framework could achieve better prediction performance than baseline methods with less training data and significantly reduce the time-consumption on model retraining.

The remaining of the paper is organised as follows: Methodology is detailed in Section~\ref{sec:method}, including the problem formulation, main framework of the proposed physics-informed deep transfer learning method, design of the generic network and the personalised network. Materials and experimental methods are presented in Section~\ref{sec:material}. Experimental results are reported in Section~\ref{sec:results}, followed by discussions in Section~\ref{sec:discussion}. Finally, conclusions are given in Section~\ref{sec:conclusion}.

\section{Methodology}
\label{sec:method}

\subsection{Problem Formulation}

In this paper, we will use sEMG measurements to predict human muscle force and joint kinematics as the exemplar. As shown in Fig.~\ref{fig:2}, after building a generic model from $EMG_t^n$ and the corresponding time steps $t$ of $m$ subjects to predict muscle forces $F^n_t$ and joint angles $\theta_t$ $\left( {n = 1, \ldots,N, t = 1, \ldots ,T} \right)$, how can we apply this model to a new subject $m+1$. Here, $N$ is the number of muscles at the joint of interest and $T$ denotes the total sample number. Since achieving the ground truths of these subject-specific data is time-consuming and labor-intensive, and it would be unwise to ignore the training data from the first $m$ subjects and extract the large amount of ground-truth data for the ($m+1$)th subject. In the following sections, we will introduce a twin physics-informed deep neural networks strategy to make the generic model personalised for the ($m+1$)th subject with limited training data from him/her.

\subsection{Main Framework of Twin Physics-informed Deep Neural Networks for Personalised Musculoskeletal Modelling}
\begin{figure*}
\centering
\includegraphics[width=1\linewidth]{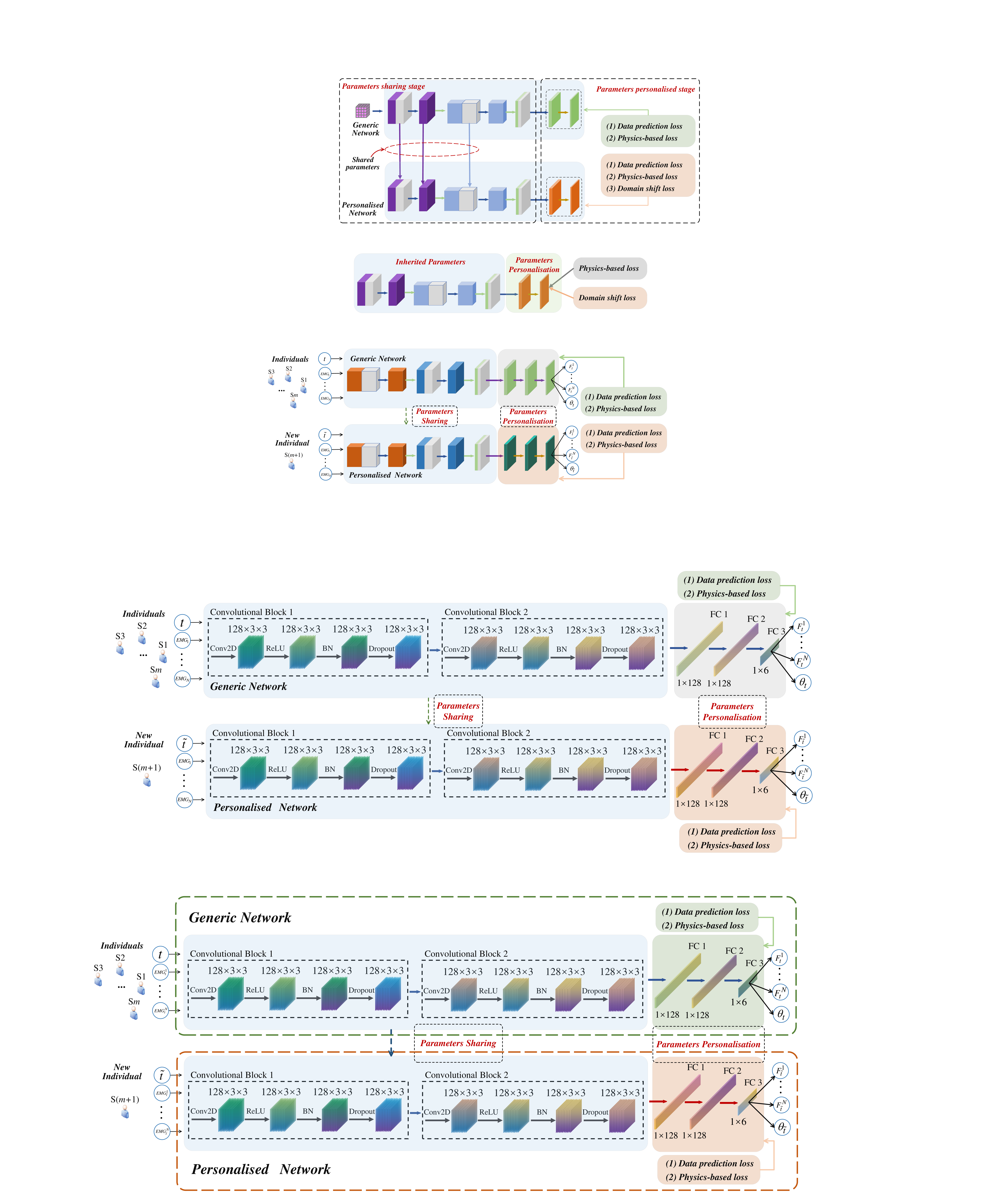}
\caption{Main framework of the proposed physics-informed deep transfer learning method. In this study, inputs of the generic network are sEMG measurements and time steps of $m$ subjects, while its outputs are muscle forces $F^n_t$ and joint angles $\theta_t$ $\left( {n = 1, \ldots,N, t = 1, \ldots ,T} \right)$. Inputs of the personalised network are sEMG measurements and the corresponding time steps of the ($m$+1)th subject, and its outputs are muscle forces $F_{\tilde t}^n$ and joint angles ${\theta _{\tilde t}}$ $\left( {\tilde t = 1, \ldots ,\tilde T} \right)$ of the subject. $n$ denotes the number of muscles at the joint of interest, $t$ and $\tilde t$ are the time steps of the $m$ subjects and the ($m$+1)th subject, respectively. We use the difference of background color of each component to demonstrate the parameter sharing and parameter personalisation phases.}
\label{fig:2}
\end{figure*}

Fig.~\ref{fig:2} depicts the main framework of the proposed physics-informed deep transfer learning method for musculoskeletal modelling, in the context of muscle forces and joint kinematics estimation from sEMG. As shown in Fig.~\ref{fig:2}, it consists of two deep neural networks, including a physics-informed generic network and a physics-informed personalised network. 

In the generic network, CNN is employed to automatically extract more discriminative features to build mappings between sEMGs and joint motions/muscle forces. Specifically, the measured sEMGs from several subjects and the corresponding time steps are first fed into CNN, and then the predicted joint angles and muscle forces are achieved with the extracted features. Such predictions should also satisfy the physical equation of motion, which is then taken as the soft constraint to penalise/regularise the loss function of CNN. In this manner, the modified loss function of the physics-informed generic network jointly minimises two components, i.e., the conventional mean square error (MSE) loss and the physics-based loss. When the new measured sEMGs of another subject are available, the parameters relating to the generic information of subjects, i.e., the parameters in the feature extraction phase, in the generic network will be directly shared to the personalised network as the backbone, which could help the personalised network hold mappings between kinematic measurements, and internal forces and muscle activations, and also guarantee the generalization. After that, the parameters relating to the subject-specific information, e.g., the parameters in fully-connected layers, in the personalised network could be achieved through fine-tuning the corresponding parameters in the generic network with the limited subject-specific data by jointly considering the data prediction error and the physical constraint. 

In this manner, the derived personalised network will contain mappings between kinematic measurements, and internal forces and muscle activations with the shared parameters from the generic network. More importantly, with many parameters frozen, over-fitting problem can be avoided during the knowledge transfer phase, and it also can significantly reduce the number of required training data and time consumption.

\subsection{Design of Generic Network}
We detail the network architecture and the modified loss function of the generic network, respectively. 

\subsubsection{Architecture of Generic Network}

The generic network is a CNN with two convolutional blocks, two fully-connected blocks, and one regression block, i.e., FC 3 in Fig.~\ref{fig:2}. To be specific, each convolutional block consists of one convolutional layer, one ReLU layer, one batch normalisation layer, and one dropout layer. In the convolutional block, the kernel size is 3, the boundary padding is 3, and the stride is 1. There are 128 kernels in the convolutional layer and a ReLU layer is added subsequently to the convolutional layer. The batch normalisation layer is employed for mitigating alternation made by the convolutonal layer. Each fully-connected block consists of one ReLU layer, one normalisation layer, and one dropout layer. The number of hidden nodes is 128. Outputs of the second fully-connected block are then fed into the regression block for the muscle forces and joint kinematics estimation.
  
\subsubsection{Loss Function of Generic Network}

The loss function of the generic network consists of the MSE loss and the physics-based loss. The MSE loss is to minimise the MSE of the ground truth and prediction, while the physics-based loss preserves the physical constraints during movements:
\begin{align}
\label{eq:total_loss}
{\mathcal{L}} &= \ {\mathcal{L}_d} + {\mathcal{L}_p} \\
{\mathcal{L}_d} &= \ MSE_{gen}\left( F \right) + MSE_{gen}\left( \theta  \right)\\
{\mathcal{L}_p} &= \ \Theta \left( {F,\theta } \right)
\end{align}
where $\mathcal{L}_d$ represents the data prediction loss of the muscle force and the joint angle, $\mathcal{L}_d$ is the loss function imposed by the physics law, which could be utilised to penalise/regularise the generic network for performance enhancement. $\Theta \left( {F,\theta } \right)$ denotes the function of predicted variables.

The MSE loss can be represented as
\begin{align}\label{eq:mse_loss}
{\ MSE_{gen}}(F) &=  \ \frac{1}{T} \displaystyle \sum_{t=1}^{T} \sum_{n=1}^{N}(F^n_t - \hat{F}^n_t)^2
\\
{\ MSE_{gen}}(\theta) &=  \frac{1}{T} \displaystyle \sum_{t=1}^{T} (\theta_t - \hat{\theta}_t)^2
\end{align}
where $F^n_t$ and $\theta_t$ are the force of muscle $n$ and the joint angle at time step $t$, and $\hat{F}^n_t$ and $\hat{\theta}_t$ denote the corresponding predictions, respectively. Furthermore, $T$ is the total sample number of subjects, and $N$ is the number of muscles at the joint of interest.

The equation of motion, which could reflect underlying relationships among the muscle force and kinematics, is utilised to design the physics-based loss: 
\begin{align}
\label{eq:physics_loss}
\Theta(F,\theta) =  \frac{1}{T}\sum_{t=1}^{T} (M(\theta_t) \Ddot{\theta}_t + C(\theta_t, \Dot{\theta}_t)+ G(\theta_t)  - \tau_t )^2
\end{align}
where $M(\theta_t), C(\theta_t, \Dot{\theta}_t)$, $G(\theta_t)$ and $\theta_t$ are the mass matrix, the Centrifugal and Coriolis force, the gravity, and the joint angle, respectively.
Additionally, $\tau_t$ is the joint torque, which can be calculated through the summation of the product of the moment arm and muscle force:
\begin{align}
\label{eq:torque}
\tau_t = \sum_{n=1}^{N} r_{n}F^n_t
\end{align}
where $r_{n}$ is the moment arm of the muscle $n$, which is exported from OpenSim.

\subsection{Design of Personalised Network}
The personalised network has the similar architecture with the generic network, thus we only demonstrate its modified loss function and the knowledge transfer process for musculoskeletal model personalisation. 

\subsubsection{Loss Function of Personalised Network}

The loss function of the personalised network also consists of the MSE loss and the physics-based loss:
\begin{align}
\label{eq:total_loss}
\tilde {\mathcal{L}} &= {{\tilde {\mathcal{L}}}_d} + {{\tilde {\mathcal{L}}}_p}\\
{{\tilde {\mathcal{L}}}_d} &= MSE_{per}\left( F \right) + MSE_{per}\left( \theta  \right)\\
{{\tilde {\mathcal{L}}}_p} &= \ \Xi \left( {F,\theta } \right)
\end{align}
where ${\tilde {\mathcal{L}}}_d$ represents the prediction loss of the muscle force and the joint angle,
${\tilde {\mathcal{L}}}_p$ is the loss function imposed by the physics law.

Specifically, the MSE loss is
\begin{align}\label{eq:mse_loss}
{\ MSE_{per}}(F) &=  \ \frac{1}{\tilde T} \displaystyle \sum_{\tilde t=1}^{\tilde T} \sum_{n=1}^{N}(F^n_{\tilde t} - \hat{F}^n_{\tilde t})^2
 \\
{\ MSE_{per}}(\theta) &=  \frac{1}{\tilde T} \displaystyle \sum_{\tilde t=1}^{\tilde T} (\theta_{\tilde t} - \hat{\theta}_{\tilde t})^2
\end{align}
where $F^n_{\tilde t}$ and $\theta_{\tilde t}$ are the force of the muscle $n$ and the joint angle at time step $\tilde t$, and $\hat{F}^n_{\tilde t}$ and $\hat{\theta}_{\tilde t}$ denote the corresponding predictions, respectively. Furthermore, $\tilde T$ is the total sample number of the new subject.

Similar to the loss function of the generic network, the equation of motion is also employed as the physics-based loss of the personalised network: 
\begin{align}
\label{eq:physics_loss}
\Xi(F,\theta) =  \frac{1}{\tilde T}\sum_{\tilde t=1}^{\tilde T} (M(\theta_{\tilde t}) \Ddot{\theta}_{\tilde t} + C(\theta_{\tilde t}, \Dot{\theta}_{\tilde t})+ G(\theta_{\tilde t})  - \tau_{\tilde t} )^2
\end{align}
where $M(\theta_{\tilde t}), C(\theta_{\tilde t}, \Dot{\theta}_{\tilde t})$, $G(\theta_{\tilde t})$ and $\theta_{\tilde t}$ are the mass matrix, the Centrifugal and Coriolis force, the gravity, and the joint angle, respectively.
Similarly, $\tau_{\tilde t}$ could be calculated by
\begin{align}
\label{eq:torque}
\tau_{\tilde t} = \sum_{n=1}^{N} r_{n}F^n_{\tilde t}
\end{align}

\subsubsection{Musculoskeletal Model Personalisation}
In order to transfer the useful knowledge to the personalised network, the parameters relating to the generic information of subjects in the generic network are first fixed and then directly shared to the personalised network, enabling the updated personalised network to hold mapping relationships between sEMG measurements, and muscle forces and joint angles of subjects. These inherited parameters could help enhance the generalization of the personalised network. In the parameters penalisation phase, optimal parameters in fully-connected layers, which actually relate to the subject-specific information, could be achieved by minimising the modified loss function only with the limited available subject-specific data as inputs of the personalised network.    

\section{Materials and Experimental Settings}
\label{sec:material}
In this section, data collection, data pre-processing, baseline methods, and evaluation criteria are detailed, respectively.

\subsection{Data Collection}
Approved by the MaPS and Engineering Joint Faculty Research Ethics Committee of University of Leeds (MEEC 18-002), {eight able-bodied subjects, including four males and four females, aged 20-30, were recruited to record data for the experiment.} All subjects gave signed consent. Specifically, subjects were informed to maintain a fully straight torso with the $90^{\circ}$ abducted shoulder and the $90^{\circ}$ flexed elbow joint. The continuous wrist flexion/extension motion was recorded by the VICON motion capture system. The joint motions were computed by the upper limb model using 16 reflective markers (sampled at~\SI{250}{Hz}). In addition, sEMGs were measured by Avanti Sensors (sampled at~\SI{2000}{Hz}) from the main wrist muscles, i.e., the $flexor~carpi~radialis$ (FCR), the $flexor~carpi~ulnaris$ (FCU), the $extensor~carpi~radialis~longus$ (ECRL), the $extensor~carpi~radialis~brevis$ (ECRB), and the $extensor~carpi~ulnaris$ (ECU), respectively. The electrodes were allocated by palpation and evaluated by performing contraction while looking at the signal before the experiment. Moreover, sEMGs and motion data were synchronised and resampled at \SI{1000}{Hz}. Five repetitive trials were done for each subject, and a three-minute break was given between trials to prevent the muscle fatigue. {We collected 80,000 samples in total (10,000 samples from each subjects).}

\subsection{Data Pre-processing}
The measured sEMGs were then band-pass filtered (\SI{20}{Hz} and~\SI{450}{Hz}), fully rectified, and low-pass filtered (\SI{6}{Hz}), respectively. After that, they were normalised concerning the maximum voluntary contraction recorded before the experiment, resulting in the enveloped EMG. The markers' data were used to compute the wrist kinematics via the inverse kinematic tool, and the joint torque and wrist muscle forces were then achieved from the inverse dynamic and the computed muscle control tools ensured the computed motion was consistent with the measured joint motion. Each wrist motion trial consists of time steps, filtered sEMGs, wrist muscle forces, and wrist joint angles, respectively.

\subsection{Baseline Methods}
To verify the effectiveness of the proposed physics-informed deep transfer learning framework, several methods are chosen as baseline methods for the comparison. In addition, we define the generic domain as sufficient labelled sEMG measurements collected from several subjects could be provided for the generic network training, denoted by ${D_G}$, while personalised domain as only limited labelled sEMG measurements collected from another specific subject are available for musculoskeletal model training or personalised network training, denoted by ${D_P}$.

\subsubsection{CNN with Training Data from $D_G$ (CNN-1)}
CNN-1 only involves conventional supervised learning with two convolutional blocks, two fully-connected blocks, and one regression block. The measured sEMGs from $D_G$ are for CNN-1 training, while sEMGs from $D_P$ for testing. Stochastic gradient descent with momentum optimiser is employed for CNN-1 training, the batch size is set as 1, the maximum iteration is set as 2000, and the initial learning rate is 0.001.

\subsubsection{CNN with Training Data from $D_P$ (CNN-2)}
CNN-2 also not involves knowledge transfer and is with the same architecture as CNN-1, but its training data are the sEMGs measured from the new subject in $D_P$. Its training strategy and parameter settings are the same as CNN-1. 

\subsubsection{CNN with Knowledge Transfer (CNN-KT)}
CNN-KT shares the same architecture, training strategy, and parameter settings as CNN-1 and CNN-2. Differently, CNN-KT is first pre-trained using the data from $D_G$, and then its parameters relating to the generic information of subjects, i.e., the parameters in the feature extraction phase, are fixed. After that, the data from the new subject are utilised for network retraining, and the parameters relating to the subject-specific information, i.e., the parameters in fully-connected layers, are achieved by fine-tuning strategy.

\subsubsection{Physics-informed CNN with Training Data from $D_G$ (Pi-CNN-1)}
Pi-CNN-1 has the same architecture and training data as CNN-1. However, its loss function is to jointly minimise the MSE loss and the physics-based loss, i.e., the loss function illustrated in Eq.(1).

\subsubsection{Physics-informed CNN with Training Data from $D_P$ (Pi-CNN-2)}
Pi-CNN-2 has the same architecture and training data as CNN-2, and its loss function is the same as the loss function of Pi-CNN-1. 

\subsection{Evaluation Criteria}
In this paper, two commonly used evaluation criteria, including root mean square error (RMSE) and Pearson's correlation coefficient ($CC$), are considered to quantify the performance of the proposed framework. To be specific, RMSE is
\begin{align}
\label{eq:nrmse}
{\rm{RMSE}} = \sqrt {\frac{1}{{\tilde T}}\sum\nolimits_{\tilde t = 1}^{\tilde T} {{{\left( {{y_{\tilde t}} - {{\hat y}_{\tilde t}}} \right)}^2}} } 
\end{align}  
where $y_{\tilde t}$ and $\hat y_{\tilde t}$ indicate the ground truth and the corresponding predicted value, respectively.

$CC$ is defined as
\begin{align}
\label{eq:r}
{{CC}} = \frac{{\sum\nolimits_{\tilde t = 1}^{\tilde T} {\left( {{y_{\tilde t}} - \overline{y_{\tilde t}}} \right)\left( {{y_{\tilde t}} - \overline{{\hat y}_{\tilde t}}} \right)} }}{{\sqrt {\sum\nolimits_{\tilde t = 1}^{\tilde T} {{{\left( {{y_{\tilde t}} - \overline{y_{\tilde t}}} \right)}^2}} } \sqrt {\sum\nolimits_{\tilde t = 1}^{\tilde T} {{{\left( {{y_{\tilde t}} - \overline{{\hat y}_{\tilde t}}} \right)}^2}} } }}
\end{align}
where $\overline{y_{\tilde t}}$ and $\overline{\hat y_{\tilde t}}$ are the mean of the ground truth and predicted value, respectively.

\section{Results}
\label{sec:results}
In this section, the effectiveness of the proposed physics-informed deep transfer learning framework for musculoskeletal modelling penalisation is verified. Specifically, the training process of the proposed framework is first shown to demonstrate its convergence. Performance evaluation for the single-to-single and the multiple-to-single scenarios are then performed to depict the predicted performance of the proposed framework. It should be noted that all the training of the proposed framework and baseline methods is carried out on a workstation with GeForce RTX 2080 Ti graphic cards and 128G RAM.

\subsection{Training Process of the Proposed Framework}
\begin{figure}
\centering
\begin{minipage}{1\linewidth}
\centering
\includegraphics[width=1.05\linewidth]{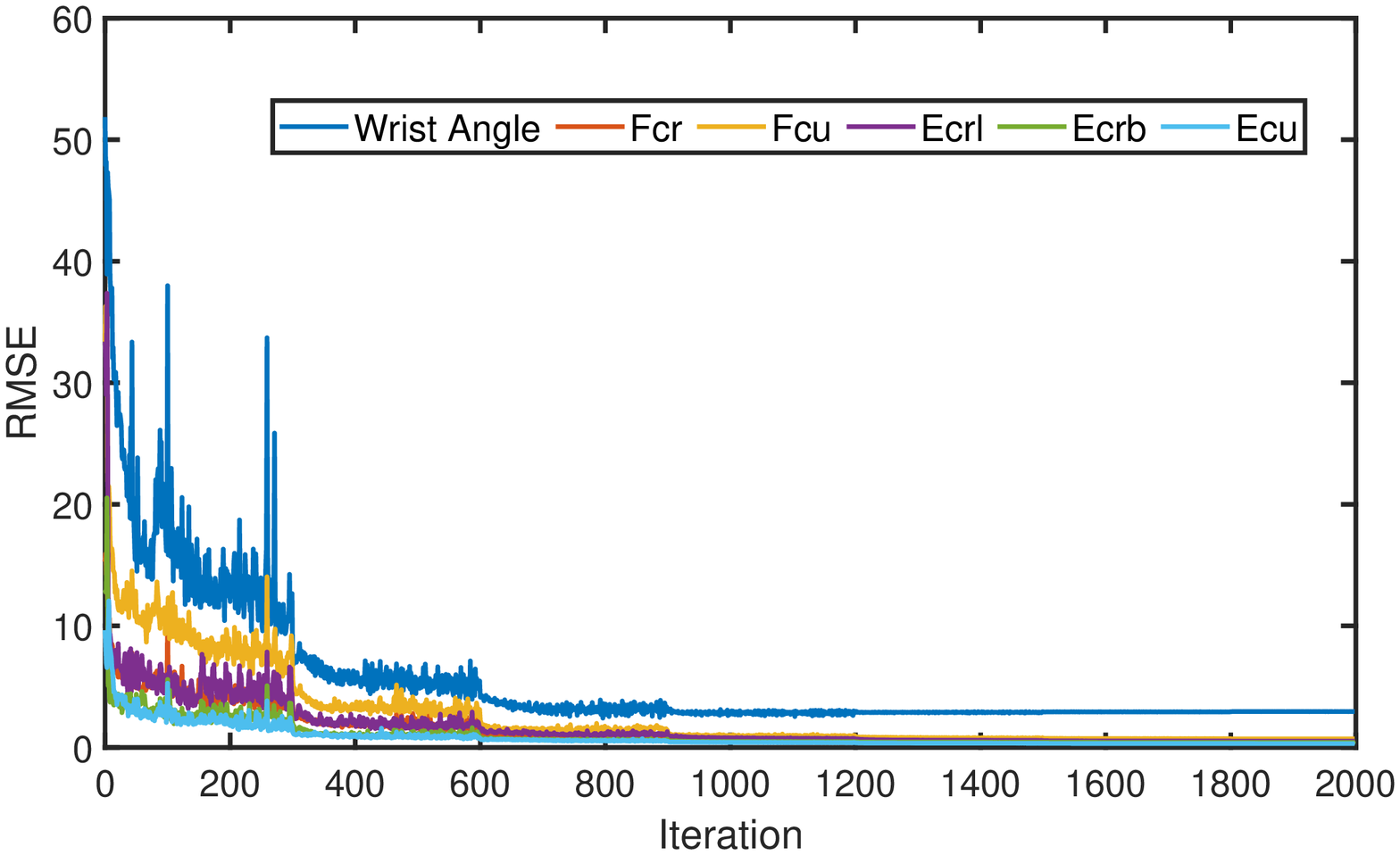}
{\fontsize{7.5pt}{9.8pt}\selectfont (a) Separate losses of Pi-CNN-1}
\end{minipage}\\
\begin{minipage}{1\linewidth}
\centering
\includegraphics[width=1.05\linewidth]{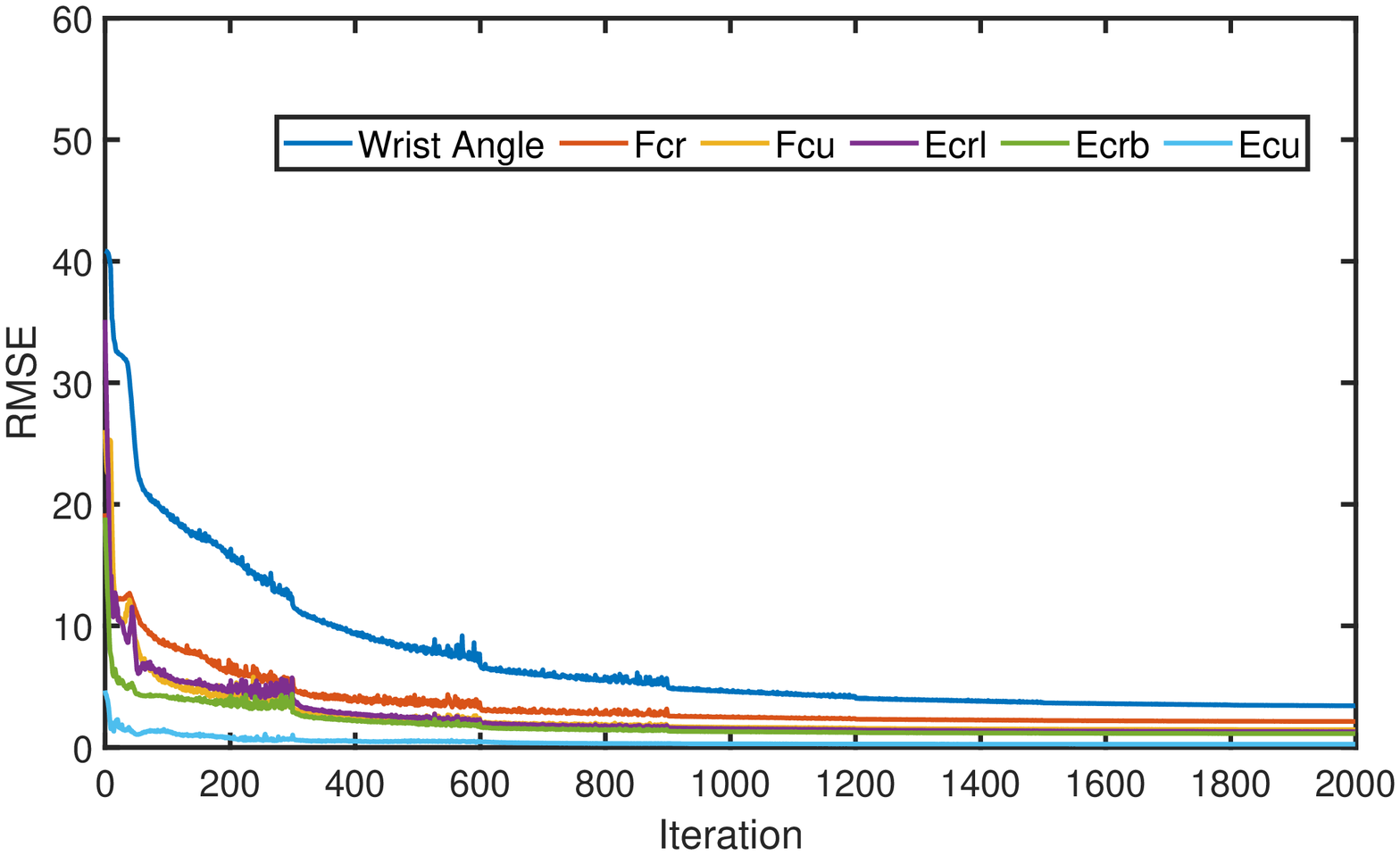}
{\fontsize{7.5pt}{9.8pt}\selectfont (b) Separate losses of Pi-CNN-KT}
\end{minipage}\\
\caption{Convergence illustration of the proposed framework. The proposed framework has faster convergence speed with less local oscillations.}
\label{fig:4}
\end{figure}

{In the experiments, we use 60$\%$ of the data to train the proposed framework and all the selected comparison methods, 20$\%$ as the validation dataset, and the rest 20$\%$ as the testing dataset. During the training phase, we set the batch size as 1, and the embedded CNN is trained by stochastic gradient descent with momentum. The maximum iteration is set as 2000, the initial learning rate is 0.001 and the dropout rate is 0.5.} The separate losses of wrist muscle forces and wrist angle estimation of Pi-CNN-1 and the proposed framework (also denoted by Pi-CNN-KT in the experiments below) during the training are depicted in Fig.~\ref{fig:4}. Observed from Fig.~\ref{fig:4}, the proposed framework has faster convergence speed than Pi-CNN-1, and its losses are more stable with less local oscillations than the ones of Pi-CNN-1. Compared with Pi-CNN-1, the twin neural networks mechanism in the proposed framework could accelerate the convergence of networks, For example, the loss of ECU in Fig.~\ref{fig:4} (b) drops significantly at the beginning of the training, and then converges.

\subsection{Performance Evaluation in Single-to-Single Scenario}
\begin{figure*}
\centering
\begin{minipage}{0.32\linewidth}
\centering
\includegraphics[width=1.1\linewidth]{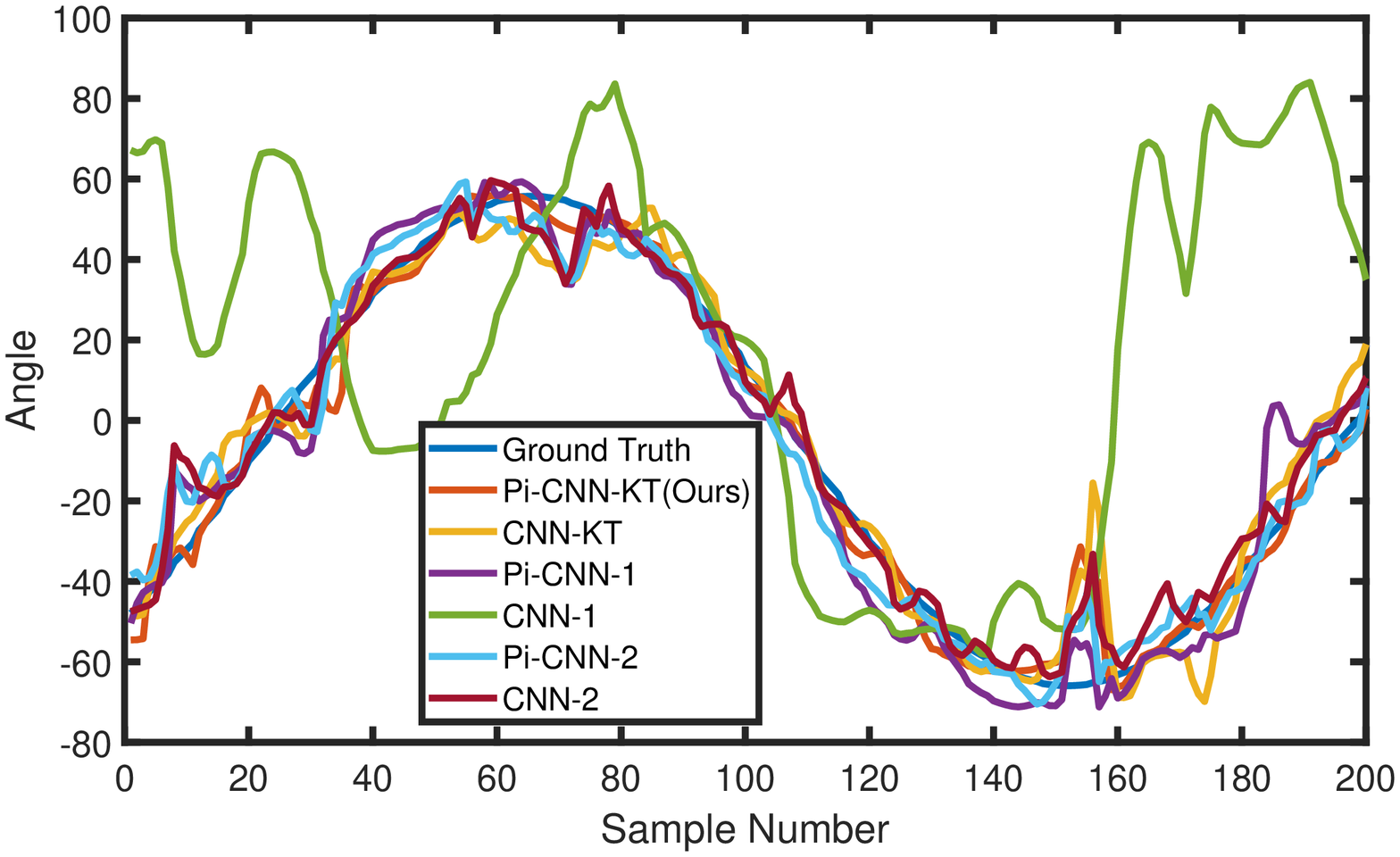}
\centerline{{\fontsize{7.5pt}{9.8pt}\selectfont (a) Predicted results of wrist angle}}
\end{minipage}
\begin{minipage}{0.32\linewidth}
\centering
\includegraphics[width=1.1\linewidth]{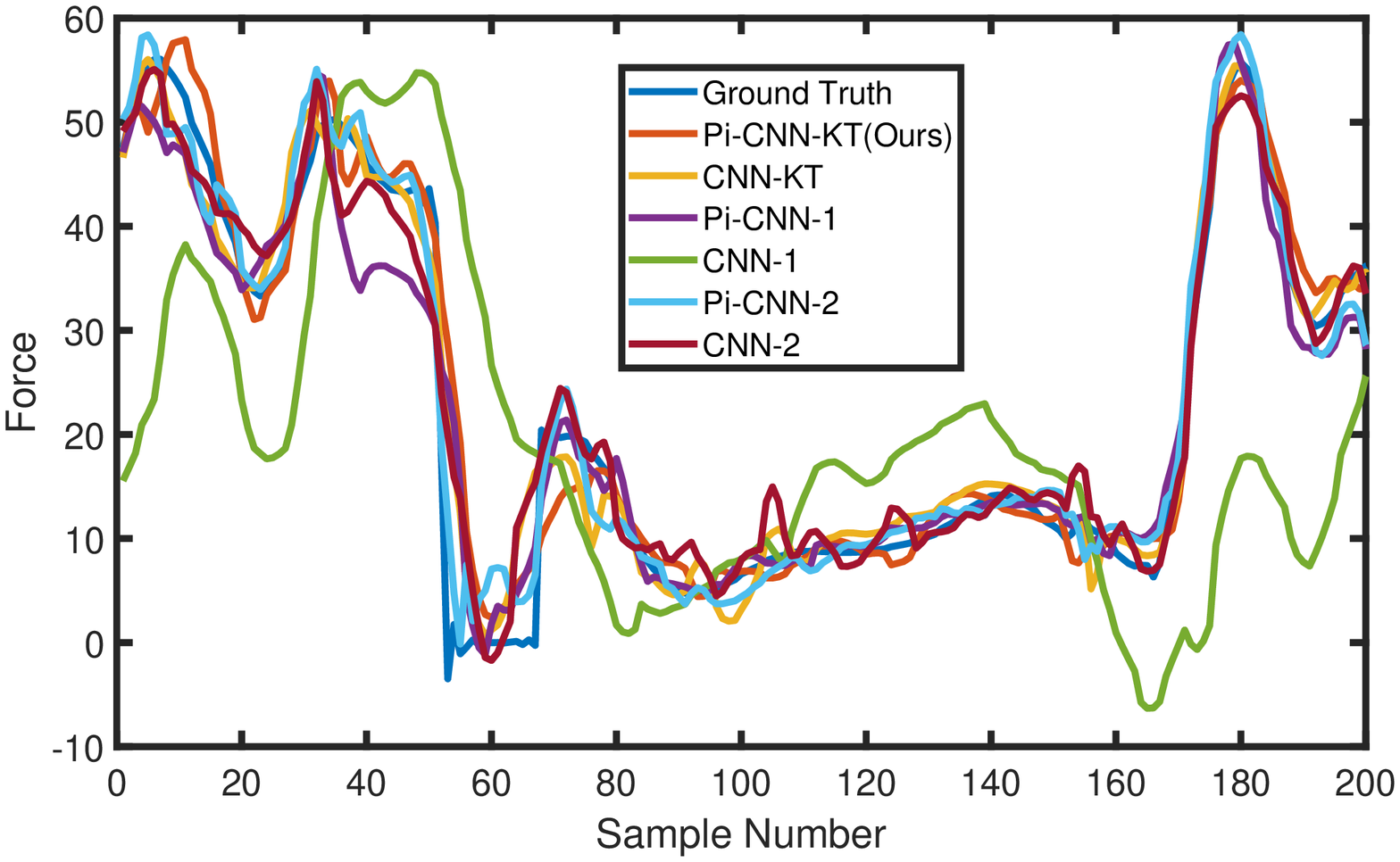}
\centerline{{\fontsize{7.5pt}{9.8pt}\selectfont (b) Predicted results of FCR}}
\end{minipage}
\begin{minipage}{0.32\linewidth}
\centering
\includegraphics[width=1.1\linewidth]{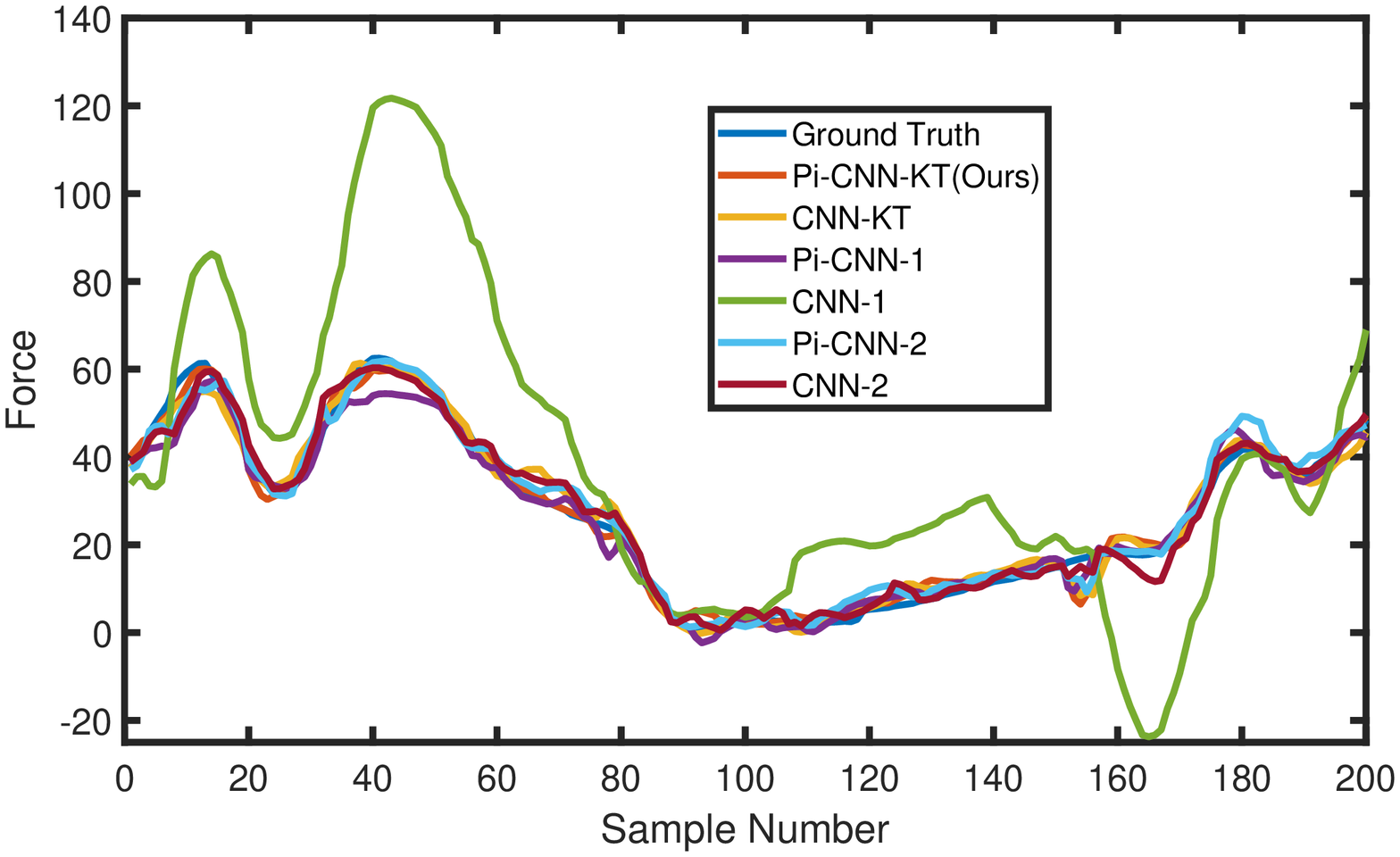}
\centerline{{\fontsize{7.5pt}{9.8pt}\selectfont (c) Predicted results of FCU}}
\end{minipage}\\
\begin{minipage}{0.32\linewidth}
\centering
\includegraphics[width=1.1\linewidth]{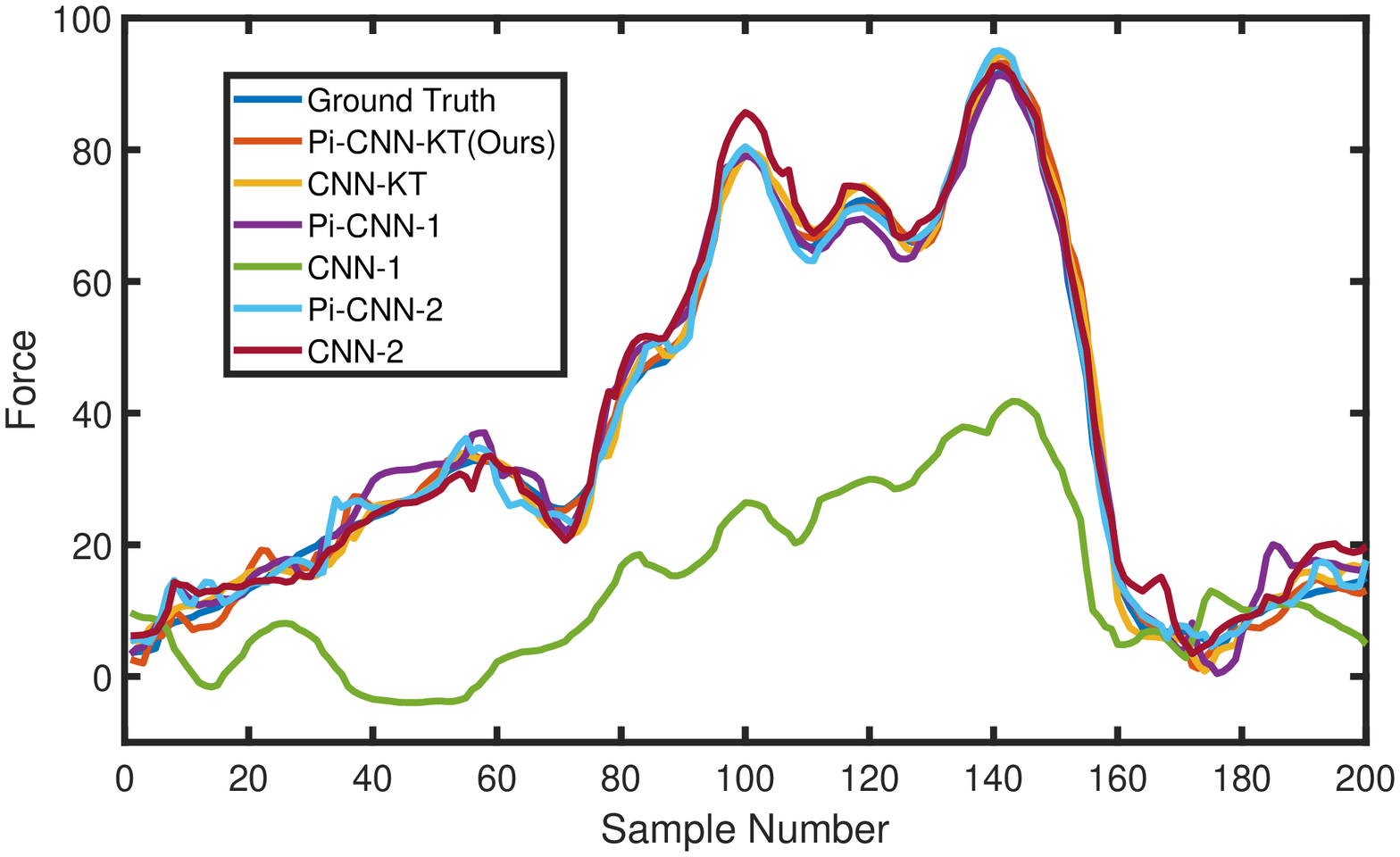}
\centerline{{\fontsize{7.5pt}{9.8pt}\selectfont (d) Predicted results of ECRL}}
\end{minipage}
\begin{minipage}{0.32\linewidth}
\centering
\includegraphics[width=1.1\linewidth]{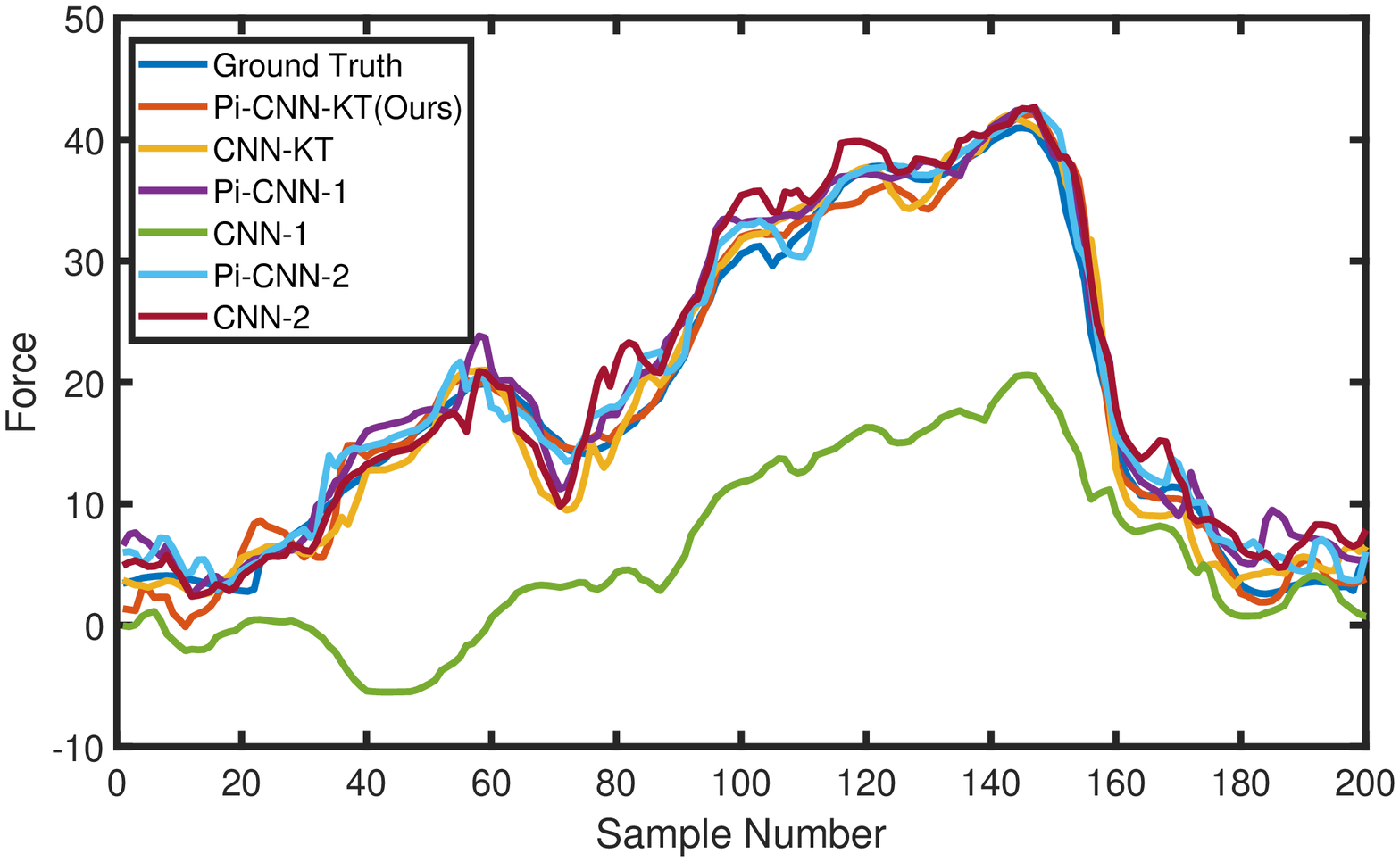}
\centerline{{\fontsize{7.5pt}{9.8pt}\selectfont (e) Predicted results of ECRB}}
\end{minipage}
\begin{minipage}{0.32\linewidth}
\centering
\includegraphics[width=1.1\linewidth]{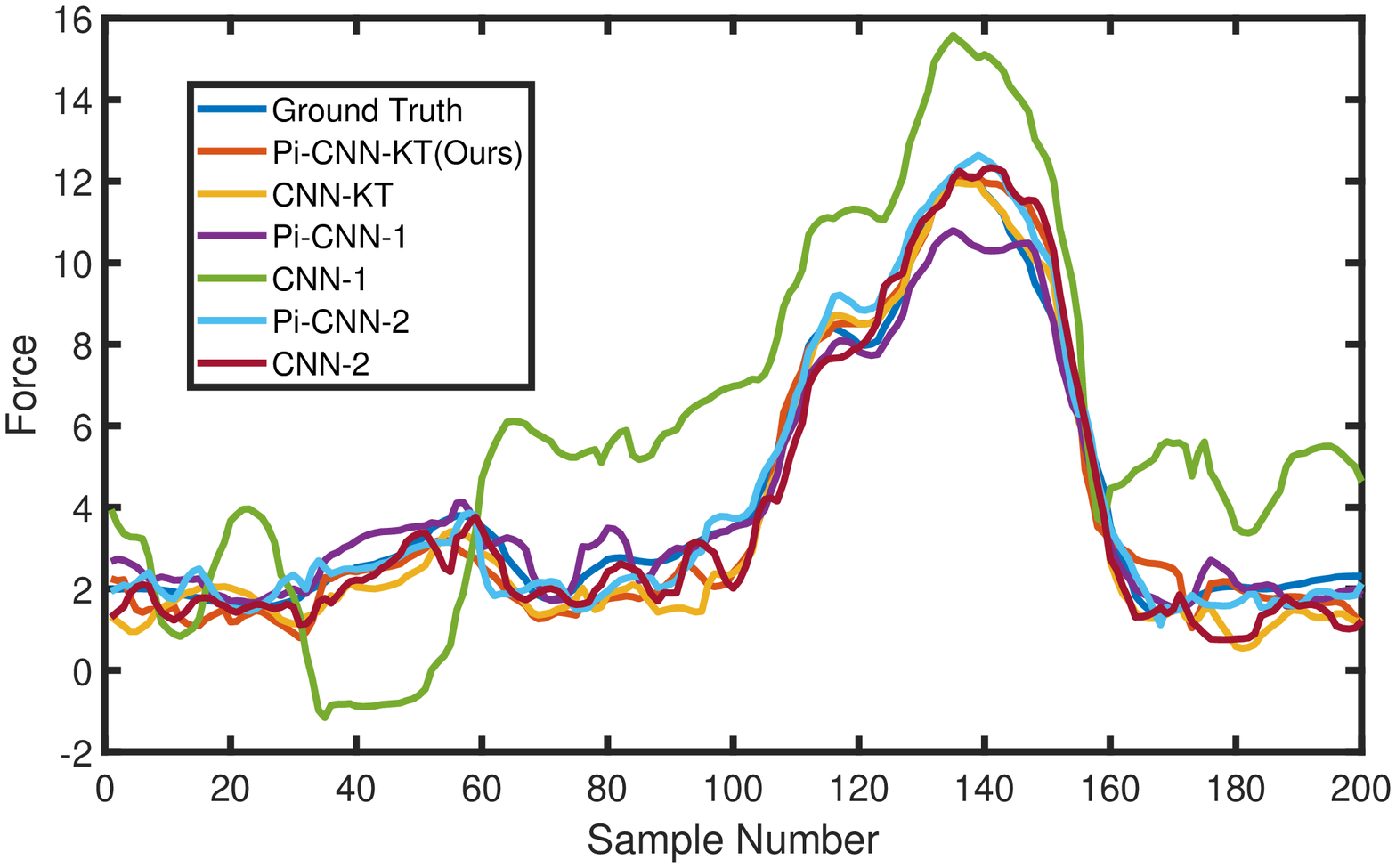}
\centerline{{\fontsize{7.5pt}{9.8pt}\selectfont (f) Predicted results of ECU}}
\end{minipage}\\
\caption{{Representative predicted results of the proposed framework and baseline methods in the single-to-single scenario. The predicted outputs are the wrist angle, muscle force of FCR, muscle force of FCU, muscle force of ECRL, muscle force of ECRB, and muscle force of ECU.}}
\label{fig:7}
\end{figure*}

We first evaluate the performance of the proposed framework in the single-to-single scenario. The generic network is first trained using the data from one specific subject, i.e., the $m$th subject ($m$=1,2,…,7), and then the personalised network is fine-tuned using the data from the 8th subject. Fig.~\ref{fig:7} illustrates the representative predicted results of the proposed framework and baseline methods, including the wrist flexion angle, muscle force of FCR, muscle force of FCU, muscle force of ECRL, muscle force of ECRB, and muscle force of ECU, respectively. According to Fig.~\ref{fig:7}, CNN-1 achieves the worst predicted results, because it does not involve the knowledge transfer of the new subject, leading to its poor generalization on the unseen data. The predicted results of CNN-KT are better than CNN-1, indicating the effectiveness of the twin neural networks mechanism employed in the musculoskeletal model personalisation. In addition, Pi-CNN-1 achieves better predicted performance than CNN-1, its means that the proposed physics-based loss function could help enhance the performance. The proposed framework achieves the best predicted performance, which indicates that the combination of the physics-based domain knowledge embedding and the twin neural networks mechanism could significantly strengthen the performance of the data-driven musculoskeletal model, and enhance its robustness and generalization.   

Table~\ref{tab:1} lists the detailed RMSEs and $CC$s of the proposed framework and baseline methods in the single-to-single scenario. Observed from Table~\ref{tab:1}, the proposed framework could achieve smaller RMSEs and higher $CC$s than baseline methods. To be specific, among the methods without physics-based domain knowledge embedding, i.e., CNN-1, CNN-2 and CNN-KT, CNN-KT has the smallest RMSEs and the highest $CC$s, it because the transferred knowledge of the new subject makes the personalised network contain subject-specific information. Moreover, the performance of CNN-2 is better than CNN-1, the main reason is that CNN-1 does not learn any statistical characteristics of the new subject, it also indicates the poor generalization of the conventional data-driven musculoskeletal model on the unseen data. Finally, we can find that the proposed framework outperforms CNN-KT, it means that the physics-based domain knowledge in the modified loss function could provide informative constraints to penalise/regularise CNN utilised in the proposed framework for performance improvement.

\begin{table*}[htbp]
    {
    \caption{Detailed RMSEs and $CC$s of the proposed framework and baseline methods in the single-to-single scenario}
    \label{tab:1}
    \centering
    \resizebox{\linewidth}{!}
    {
    \begin{tabular}{c|c|ccccccc|c|c|ccccccc}
    \toprule
    \bfseries Outputs & \bfseries Methods  & \bfseries S1 & \bfseries S2 & \bfseries S3 & \bfseries S4 & \bfseries S5 & \bfseries S6 & \bfseries S7 & \bfseries Outputs & \bfseries Methods & \bfseries S1 & \bfseries S2 & \bfseries S3 & \bfseries S4 & \bfseries S5 & \bfseries S6 & \bfseries S7\\
    \midrule
    \multirow{6}{*}{\bfseries Angle}  &  \textbf{Pi-CNN-KT} & \textbf{10.67/0.97} & \textbf{9.25/0.97} & \textbf{11.95/0.96} & \textbf{9.81/0.97} & \textbf{9.94/0.97} & \textbf{10.35/0.97} & \textbf{11.33/0.96} && \textbf{Pi-CNN-KT} & \textbf{4.45/0.95} & \textbf{4.02/0.96} & \textbf{4.17/0.96} & \textbf{3.99/0.97} & \textbf{4.21/0.96} & \textbf{4.32/0.96} & \textbf{4.02/0.96}\\
    \cmidrule{2-9} \cmidrule{11-18}
   & CNN-KT & 12.15/0.95 & 13.96/0.93 & 15.03/0.94 & 13.97/0.94 & 11.27/0.95 & 12.27/0.95 & 12.41/0.94 && CNN-KT & 4.53/0.95 & 4.41/0.95 & 4.95/0.95 & 4.27/0.96 & 4.79/0.94 & 4.61/0.95 & 4.79/0.95\\
    \cmidrule{2-9} \cmidrule{11-18}
    & Pi-CNN-1 & 16.55/0.93 & 17.80/0.93 & 15.39/0.93 & 16.53/0.93 & 14.91/0.94 & 16.27/0.93 & 15.99/0.93 & \multirow{6}{*}{\bfseries FCR} & Pi-CNN-1 & 5.89/0.92 & 4.72/0.95 & 4.39/0.96 & 5.77/0.93 & 5.72/0.92 & 5.66/0.93 & 5.70/0.93\\
    \cmidrule{2-9} \cmidrule{11-18}
    & CNN-1 & 67.21/0.31 & 59.26/0.35 & 71.22/0.29 & 70.09/0.31 & 62.37/0.32 & 76.29/0.28 & 65.52/0.32 && CNN-1 & 14.82/0.57 & 12.98/0.62 & 17.66/0.52 & 15.93/0.57 & 13.98/0.61 & 14.56/0.59& 13.87/0.61\\
    \cmidrule{2-9} \cmidrule{11-18}
    & Pi-CNN-2 & 14.95/0.94 & 16.21/0.93 & 17.98/0.93 & 15.63/0.93 & 12.96/0.95 & 14.56/0.94 & 13.29/0.94 & & Pi-CNN-2 & 4.82/0.95& 4.69/0.95 & 5.63/0.93 & 4.81/0.94 & 4.98/0.94 & 5.03/0.95& 4.91/0.95\\
    \cmidrule{2-9} \cmidrule{11-18}
    & CNN-2 & 17.24/0.93 & 19.22/0.92 & 18.31/0.93 & 18.09/0.92 & 14.15/0.94 & 16.98/0.98 & 16.90/0.93 & & CNN-2 & 5.79/0.93& 5.66/0.93 & 6.07/0.93 & 5.75/0.93 & 5.95/0.93 & 5.62/0.93& 5.53/0.94\\
   \midrule
    \multirow{6}{*}{\bfseries FCU} &  \textbf{Pi-CNN-KT} & \textbf{2.82/0.99} & \textbf{3.09/0.98} & \textbf{4.51/0.98} & \textbf{2.99/0.98} & \textbf{3.29/0.98} & \textbf{3.02/0.98} & \textbf{2.77/0.99} & & \textbf{Pi-CNN-KT} & \textbf{3.13/0.99} & \textbf{4.19/0.97} & \textbf{3.83/0.99} & \textbf{3.29/0.99} & \textbf{3.02/0.99} & \textbf{3.22/0.99} & \textbf{3.09/0.99}\\
    \cmidrule{2-9} \cmidrule{11-18}
    & CNN-KT & 3.05/0.98 & 3.82/0.98 & 5.23/0.97 & 3.93/0.98 & 3.73/0.98 & 3.72/0.98 & 3.19/0.98 & & CNN-KT & 3.61/0.99 & 4.92/0.95 & 4.20/0.97 & 3.55/0.99 & 3.94/0.98 & 3.69/0.99 & 3.67/0.99\\
    \cmidrule{2-9} \cmidrule{11-18}
    & Pi-CNN-1 & 4.21/0.98 & 4.53/0.98 & 5.89/0.96 & 4.10/0.98 & 3.96/0.98 & 4.69/0.97 & 4.11/0.98 & \multirow{6}{*}{\bfseries ECRL} & Pi-CNN-1 & 4.68/0.98 & 5.87/0.95 & 5.31/0.93 & 4.77/0.97 & 4.31/0.97 & 4.06/0.98 & 3.98/0.97\\
    \cmidrule{2-9} \cmidrule{11-18}
    & CNN-1 & 21.41/0.83 & 23.60/0.82 & 31.99/0.80 & 19.57/0.83 & 18.28/0.86 & 20.33/0.83 & 22.30/0.82 && CNN-1 & 26.81/0.85 & 30.95/0.82 & 27.61/0.86 & 29.30/0.87 & 27.90/0.86 & 24.91/0.85& 22.87/0.83\\
    \cmidrule{2-9} \cmidrule{11-18}
    & Pi-CNN-2 & 3.34/0.98 & 4.27/0.98 & 5.78/0.97 & 3.91/0.98 & 4.20/0.98 & 3.99/0.98 & 3.51/0.98 & & Pi-CNN-2 & 4.21/0.97& 5.51/0.93 & 4.72/0.97 & 4.56/0.98 & 4.07/0.98 & 3.98/0.98 & 3.92/0.98\\
    \cmidrule{2-9} \cmidrule{11-18}
    & CNN-2 & 4.36/0.98 & 4.99/0.97 & 6.31/0.96 & 4.25/0.98 & 4.07/0.98 & 4.51/0.97 & 4.27/0.98 & & CNN-2 & 4.59/0.98& 5.72/0.94 & 5.17/0.95 & 4.72/0.98 & 4.19/0.97 & 4.26/0.97& 4.05/0.97\\
    \midrule
    \multirow{6}{*}{\bfseries ECRB} &  \textbf{Pi-CNN-KT} & \textbf{2.45/0.98} & \textbf{2.65/0.98} & \textbf{2.32/0.98} & \textbf{2.29/0.98} & \textbf{2.02/0.98} & \textbf{2.36/0.97} & \textbf{2.22/0.98} && \textbf{Pi-CNN-KT} & \textbf{0.61/0.98} & \textbf{0.58/0.97} & \textbf{0.62/0.98} & \textbf{0.55/0.98} & \textbf{0.63/0.98} & \textbf{0.66/0.98} & \textbf{0.57/0.98}\\
    \cmidrule{2-9} \cmidrule{11-18}
    & CNN-KT & 2.72/0.97 & 2.89/0.97 & 2.66/0.98 & 2.98/0.97 & 2.58/0.98 & 2.77/0.98 & 2.57/0.98 & & CNN-KT & 0.71/0.98 & 0.82/0.97 & 0.76/0.98 & 0.81/0.97 & 0.75/0.97 & 0.76/0.98 & 0.69/0.98\\
    \cmidrule{2-9} \cmidrule{11-18}
    & Pi-CNN-1 & 3.65/0.96 & 3.89/0.96 & 3.26/0.97 & 3.98/0.94 & 3.55/0.96 & 3.27/0.97 & 3.39/0.96 & \multirow{6}{*}{\bfseries ECU} & Pi-CNN-1 & 0.86/0.96 & 0.99/0.97 & 0.81/0.97 & 0.92/0.97 & 0.85/0.97 & 0.79/0.97 & 0.82/0.97\\
    \cmidrule{2-9} \cmidrule{11-18}
    & CNN-1 & 16.06/0.75 & 15.97/0.75 & 12.50/0.77 & 16.28/0.73 & 13.33/0.78 & 15.99/0.75 & 14.27/0.76 && CNN-1 & 4.21/0.69 & 6.30/0.59 & 4.39/0.69 & 4.37/0.70 & 4.14/0.71 & 4.06/0.75& 4.13/0.70\\
    \cmidrule{2-9} \cmidrule{11-18}
    & Pi-CNN-2 & 3.12/0.97 & 3.57/0.96 & 3.19/0.97 & 3.55/0.96 & 3.05/0.98 & 3.11/0.98 & 2.98/0.98 & & Pi-CNN-2 & 0.62/0.98& 0.80/0.97 & 0.71/0.98 & 0.85/0.97 & 0.76/0.98 & 0.72/0.98& 0.75/0.98\\
    \cmidrule{2-9} \cmidrule{11-18}
    & CNN-2 & 3.29/0.97 & 4.01/0.95 & 3.21/0.97 & 4.26/0.95 & 3.79/0.97 & 3.50/0.96 & 3.20/0.97 & & CNN-2 & 0.93/0.96& 1.16/0.93 & 0.99/0.97 & 0.91/0.96 & 0.82/0.97 & 0.80/0.98& 0.86/0.98\\
    \bottomrule
    \end{tabular}}}
\end{table*}

To further verify the feasibility of the proposed framework, a pairwise analysis between the proposed method and each comparison method is considered. One-way analysis of variance (ANOVA) is conducted for statistical analysis of the proposed framework and baseline methods, where RMSE is the response variable. A \textit{post-hoc} analysis using Tukey’s Honest Significant Difference test is applied. The significance level is set at $p < 0.05$. Fig. 4 and Fig. 5 depict the RMSEs and $CC$s of the proposed framework and baseline methods, we can find that the proposed framework achieves better predicted performance no matter which subject's data is employed to train the generic network. Table II details the multiple comparison results in terms of RMSEs and $CC$s. The multiple comparison correction method is Dunnett’s test. The proposed framework is regarded as the control group. The comparison results indicate that the proposed framework is statistically superior to baseline methods.

\begin{figure}
{
\centering
\includegraphics[width=1\linewidth]{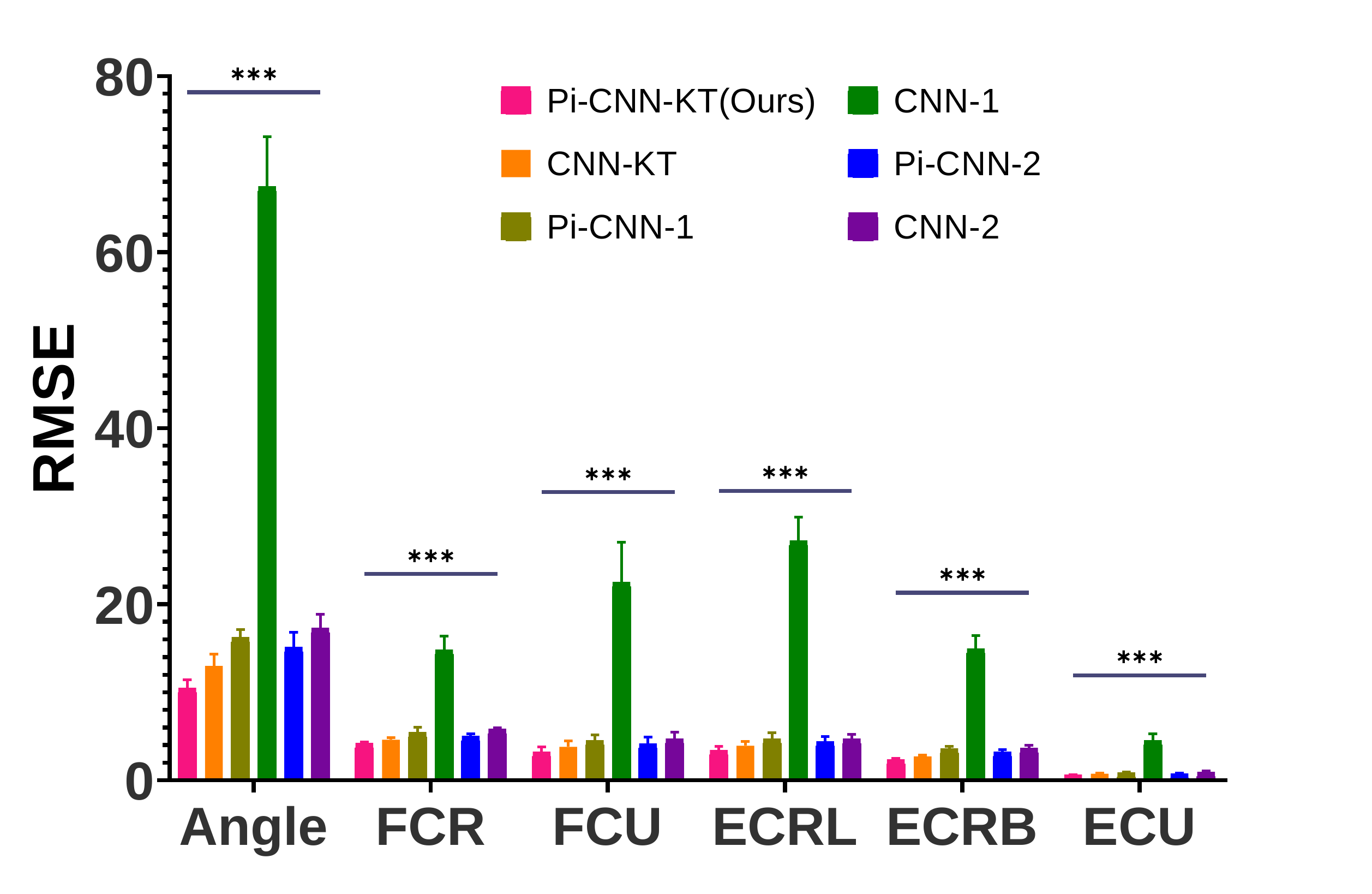}
\caption{RMSEs of the proposed framework and baseline methods with different subjects' data as the training samples for the generic network in the single-to-single scenario. The proposed framework achieves smaller RMSEs than baseline methods. The significance level is set as 0.05 ($^{***}p < 0.001,~^{**}p < 0.01, and~^{*}p < 0.05)$.}}
\label{fig:18}
\end{figure}

\begin{table}
    {
    \label{tab:my_label}
    \caption{Multiple comparison on RMSE between the proposed framework and baseline methods in the single-to-single scenario}
    \centering
    \resizebox{\linewidth}{!}{
    \begin{tabular}{c|c|c|c|c|c|c|c}
    \toprule
    \multicolumn{8}{c}{\bfseries RMSE}   \\
    \midrule[1pt]
    \multicolumn{2}{c|}{\bfseries Methods} & \bfseries Angle & \bfseries FCR & \bfseries FCU & \bfseries ECRL & \bfseries ECRB & \bfseries ECU \\
    \midrule
    CNN-1 & \textbf{Pi-CNN-KT} & $ < 0.001$ &$ < 0.001$  &$ < 0.001$  & $ < 0.001$ &$ < 0.001$ & $ < 0.001$ \\ 
    \midrule
    Pi-CNN-1 & \textbf{Pi-CNN-KT} &$ < 0.01$ & $ < 0.05$&$ 0.620$ & $ 0.176$& $ < 0.01$ & $ 0.458$  \\ 
    \midrule
    CNN-KT & \textbf{Pi-CNN-KT} &$ 0.270$ &$ 0.633$ &$ 0.969$  & $ 0.870$& $0.686$ & $ 0.854$ \\ 
    \midrule
    Pi-CNN-2 & \textbf{Pi-CNN-KT} & $ < 0.05$ & $ 0.137$&$ 0.844$ & $ 0.390$& $0.071$ & $ 0.891$ \\
    \midrule
    CNN-2 & \textbf{Pi-CNN-KT} &$< 0.001$&$ < 0.01$ & $0.5$ & $0.198$& $ < 0.01$ & $ 0.270$ \\
    \midrule[1pt]
    \multicolumn{8}{c}{\bfseries $CC$}\\
    \midrule[1pt]
    \multicolumn{2}{c|}{\bfseries Methods} & \bfseries Angle & \bfseries FCR & \bfseries FCU & \bfseries ECRL & \bfseries ECRB & \bfseries ECU \\
    \midrule
    CNN-1 & \textbf{Pi-CNN-KT} & $ < 0.001$ &$ < 0.001$  &$ < 0.001$  & $ < 0.001$ &$ < 0.001$ & $ < 0.001$ \\ 
    \midrule
    Pi-CNN-1 & \textbf{Pi-CNN-KT} &$<0.001$ & $<0.05$ & $0.457$ & $<0.05$ & $ < 0.05$ & $ 0.852$  \\ 
    \midrule
    CNN-KT & \textbf{Pi-CNN-KT} &$<0.01$&$ 0.667$ &$ 0.848$  & $ 0.865$& $0.999$ & $ 0.999$ \\ 
    \midrule
    Pi-CNN-2 & \textbf{Pi-CNN-KT} &$<0.01$&$ 0.261$ &$ 0.848$  & $ 0.177$& $0.720$ & $ 1$ \\ 
    \midrule
    CNN-2 & \textbf{Pi-CNN-KT} &$< 0.001$&$ < 0.01$ & $0.291$ & $0.062$& $ < 0.05$ & $ 0.778$ \\
    \bottomrule
    \end{tabular}}}
\end{table}

\begin{figure}
{
\centering
\includegraphics[width=1\linewidth]{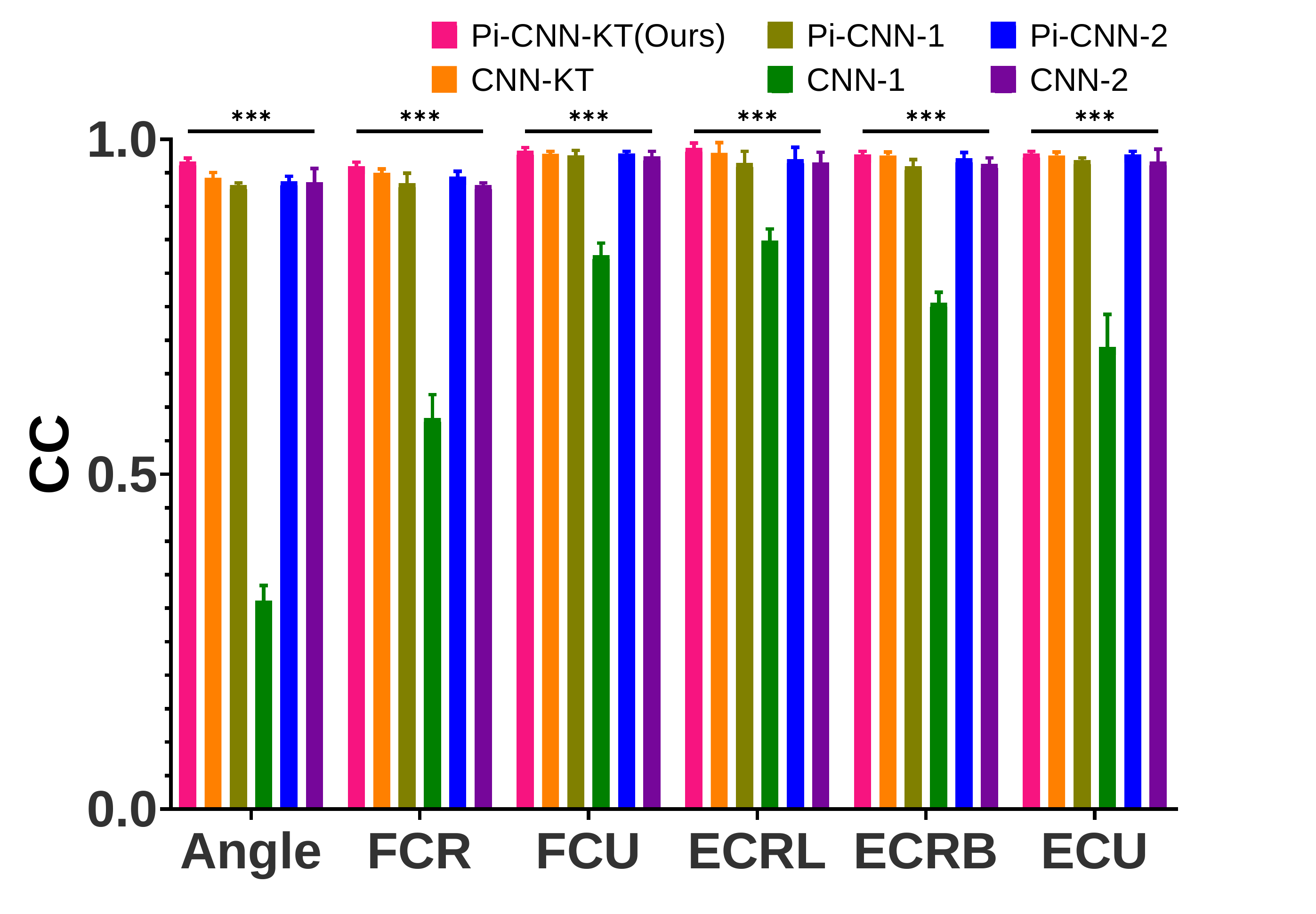}
\caption{$CC$s of the proposed framework and baseline methods with different subjects' data as the training samples for the generic network in the single-to-single scenario. The proposed framework achieves higher $CC$s than baseline methods. The significance level is set as 0.05 ($^{***}p < 0.001,~^{**}p < 0.01, and~^{*}p < 0.05)$.}}
\label{fig:19}
\end{figure}

\subsection{Performance Evaluation in Multiple-to-Single Scenario}
Aside from the single-to-single scenario, we further evaluate the performance of the proposed framework in the multiple-to-single scenario. Specifically, we first randomly choose seven subjects from the eight subjects, and the generic network is then trained using the data from the chosen seven subjects, and the data from the rest one subject is utilised to fine-tune the personalised network.

\begin{figure*}
\centering
\begin{minipage}{0.32\linewidth}
\centering
\includegraphics[width=1.1\linewidth]{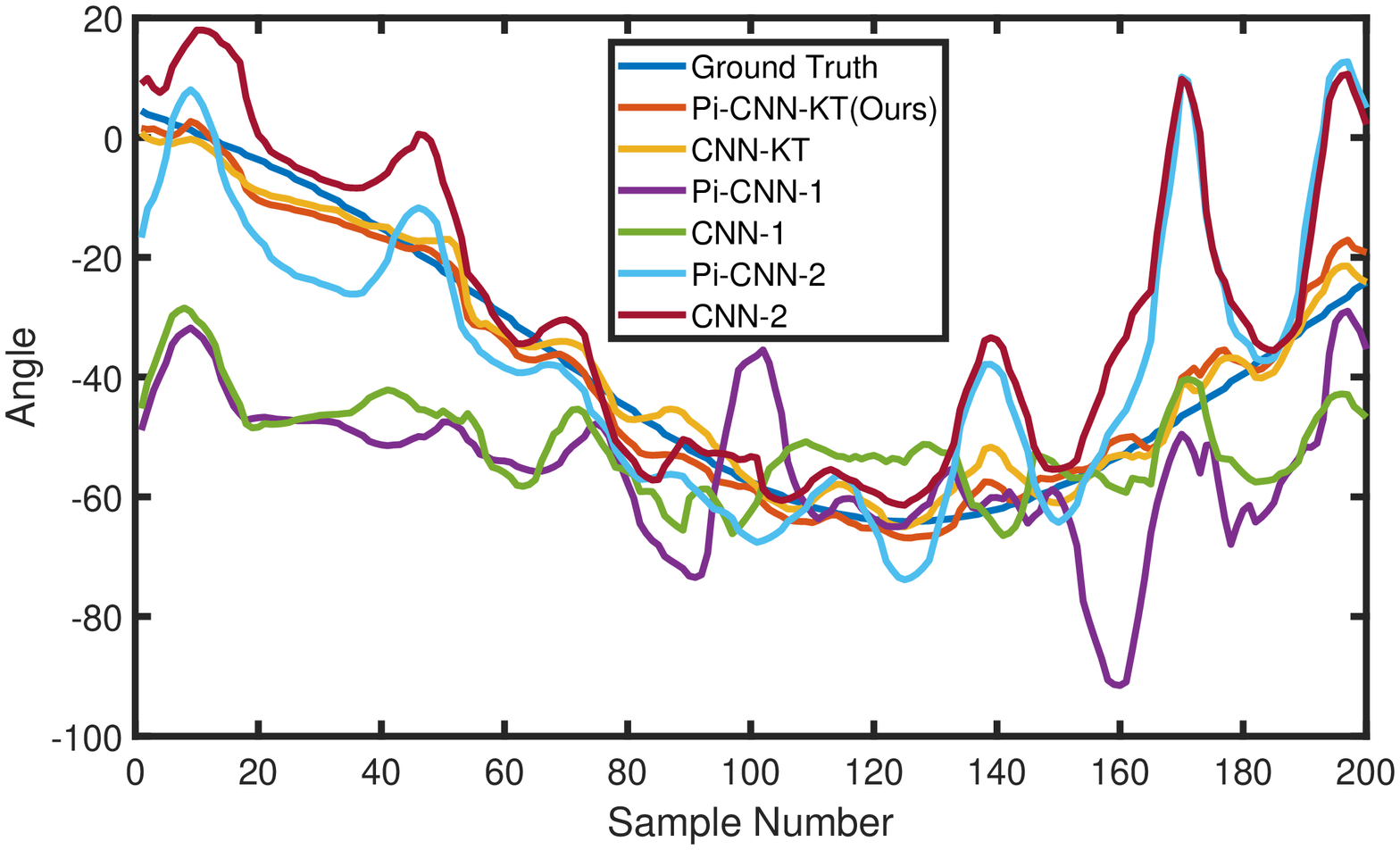}
\centerline{{\fontsize{7.5pt}{9.8pt}\selectfont (a) Predicted results of wrist angle}}
\end{minipage}
\begin{minipage}{0.32\linewidth}
\centering
\includegraphics[width=1.1\linewidth]{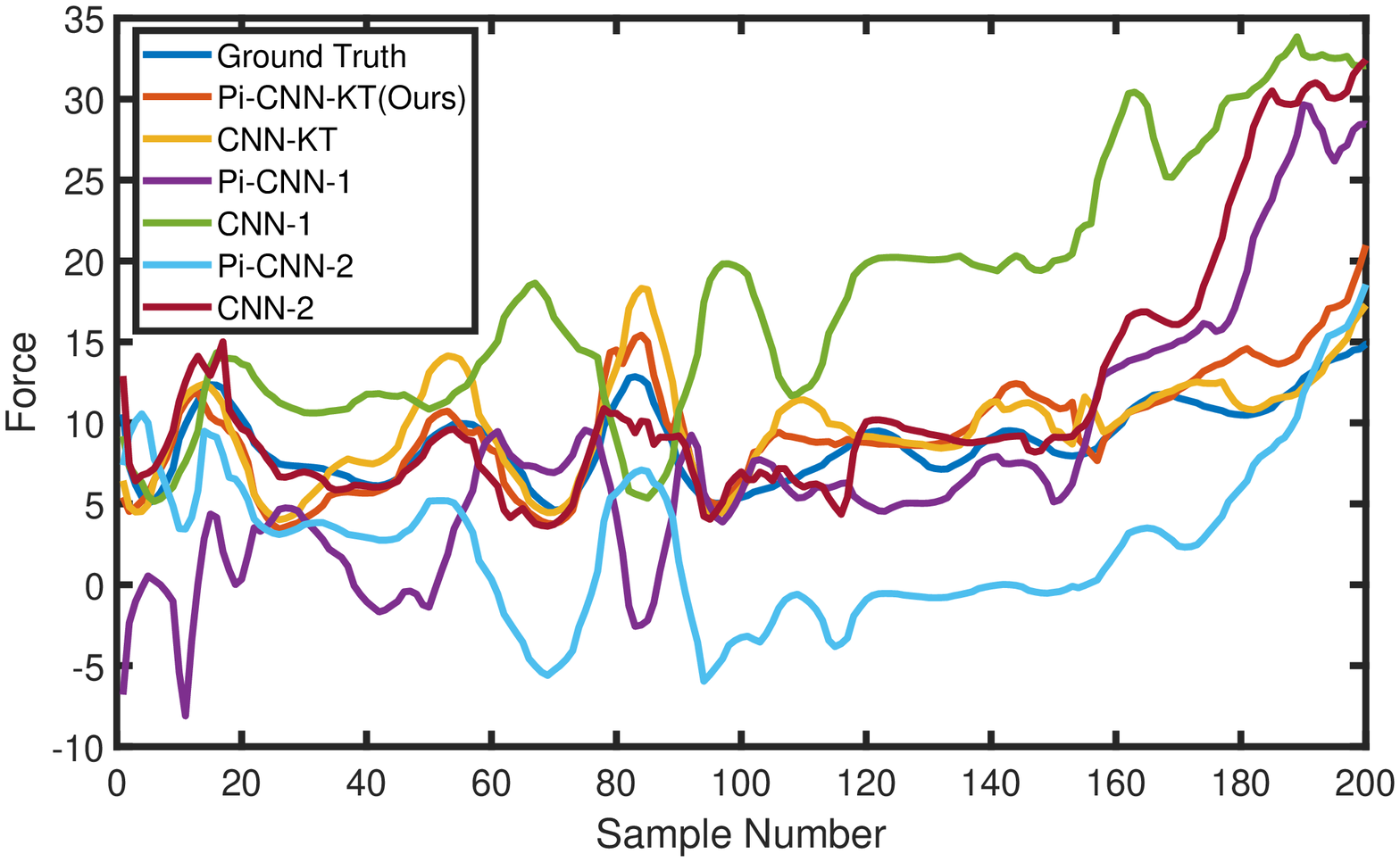}
\centerline{{\fontsize{7.5pt}{9.8pt}\selectfont (b) Predicted results of FCR}}
\end{minipage}
\begin{minipage}{0.32\linewidth}
\centering
\includegraphics[width=1.1\linewidth]{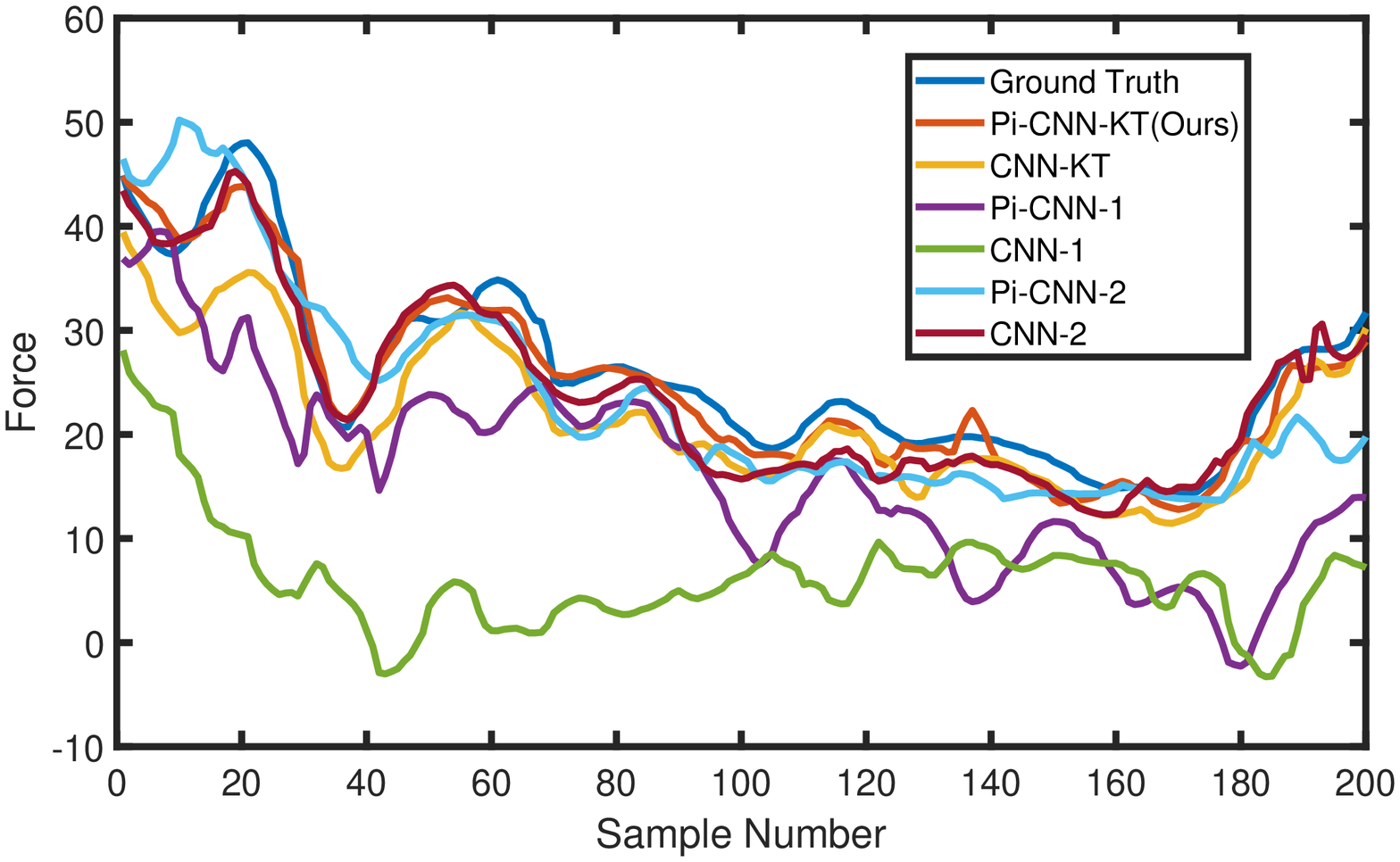}
\centerline{{\fontsize{7.5pt}{9.8pt}\selectfont (c) Predicted results of FCU}}
\end{minipage}\\
\begin{minipage}{0.32\linewidth}
\centering
\includegraphics[width=1.1\linewidth]{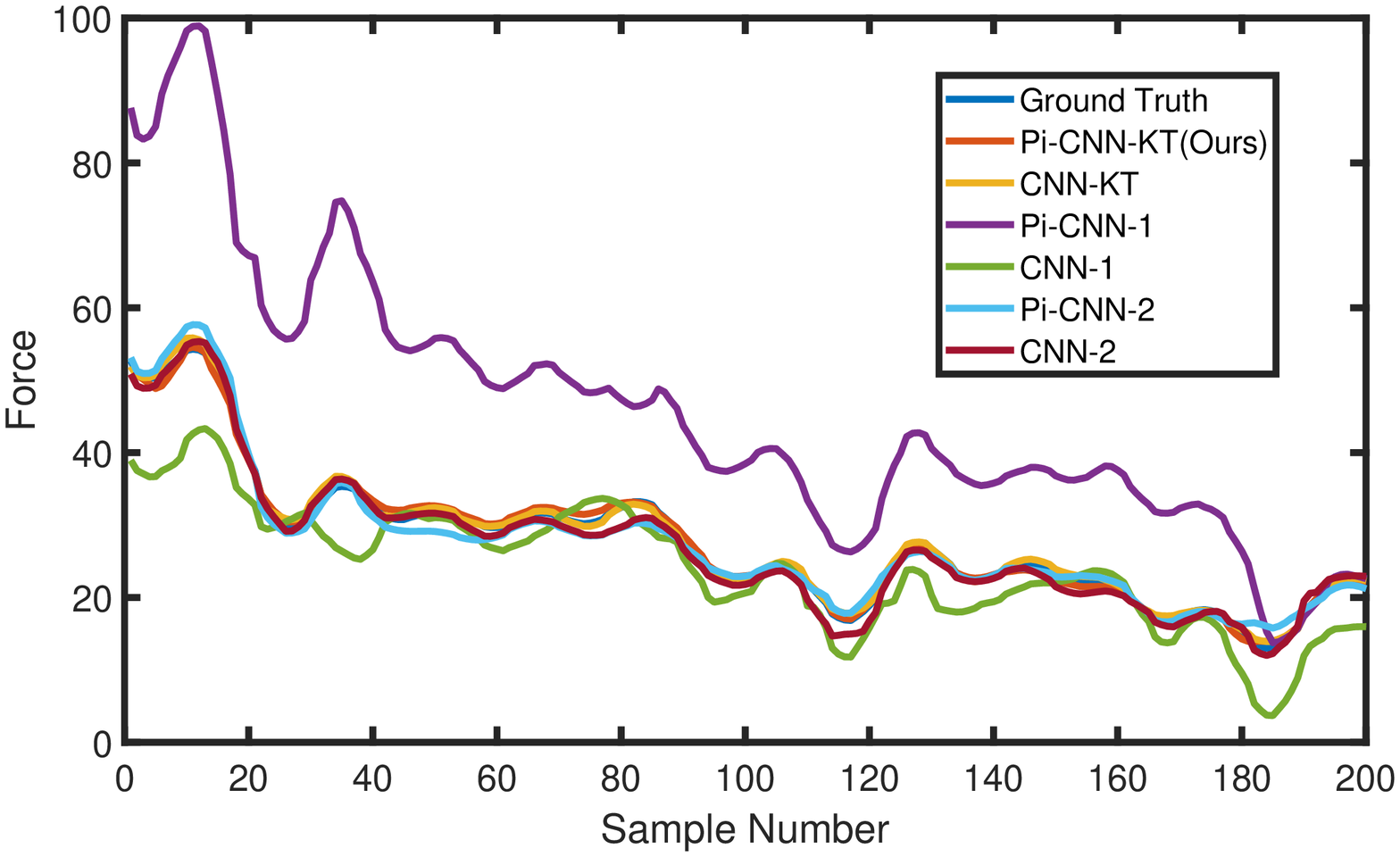}
\centerline{{\fontsize{7.5pt}{9.8pt}\selectfont (d) Predicted results of ECRL}}
\end{minipage}
\begin{minipage}{0.32\linewidth}
\centering
\includegraphics[width=1.1\linewidth]{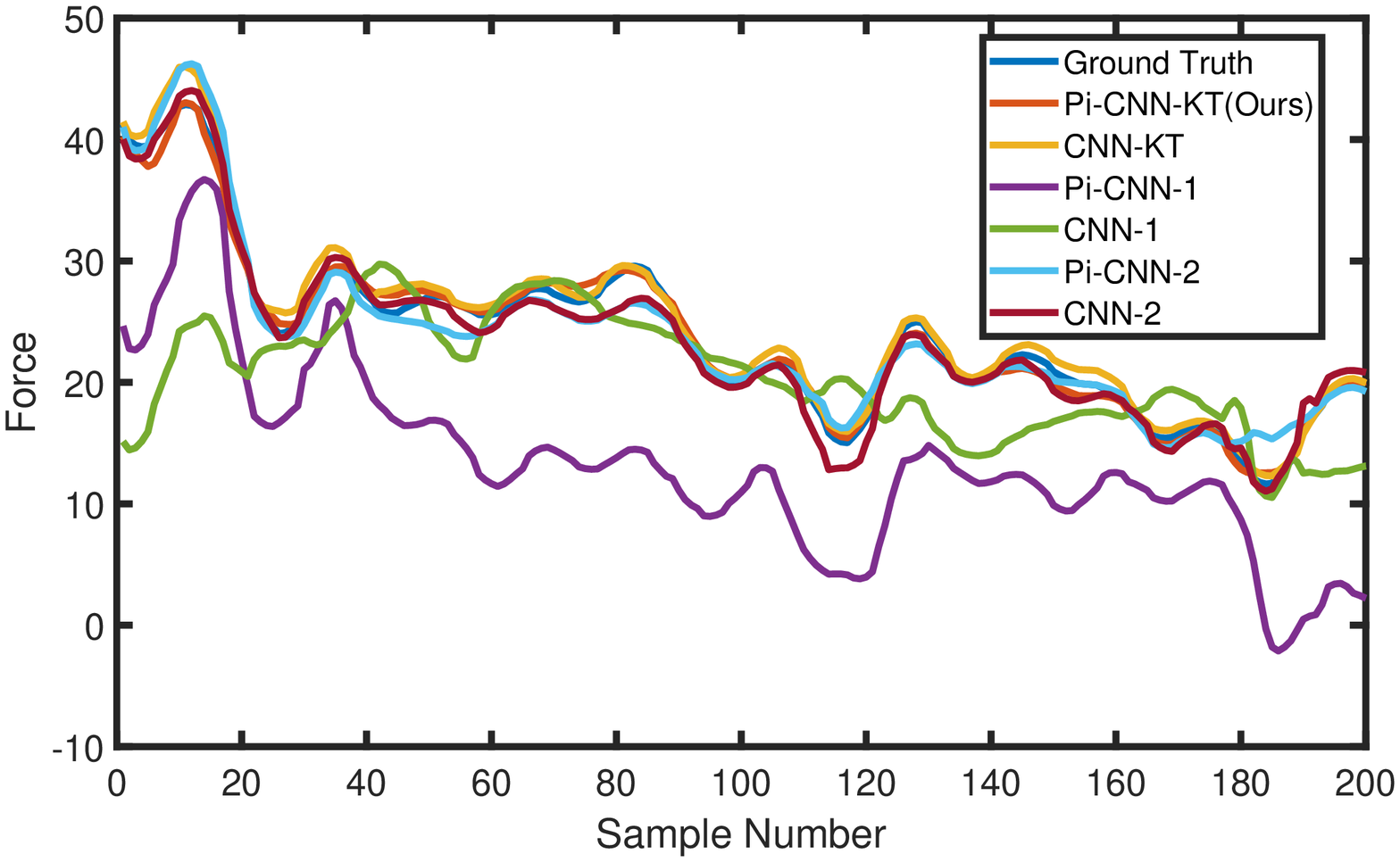}
\centerline{{\fontsize{7.5pt}{9.8pt}\selectfont (e) Predicted results of ECRB}}
\end{minipage}
\begin{minipage}{0.32\linewidth}
\centering
\includegraphics[width=1.1\linewidth]{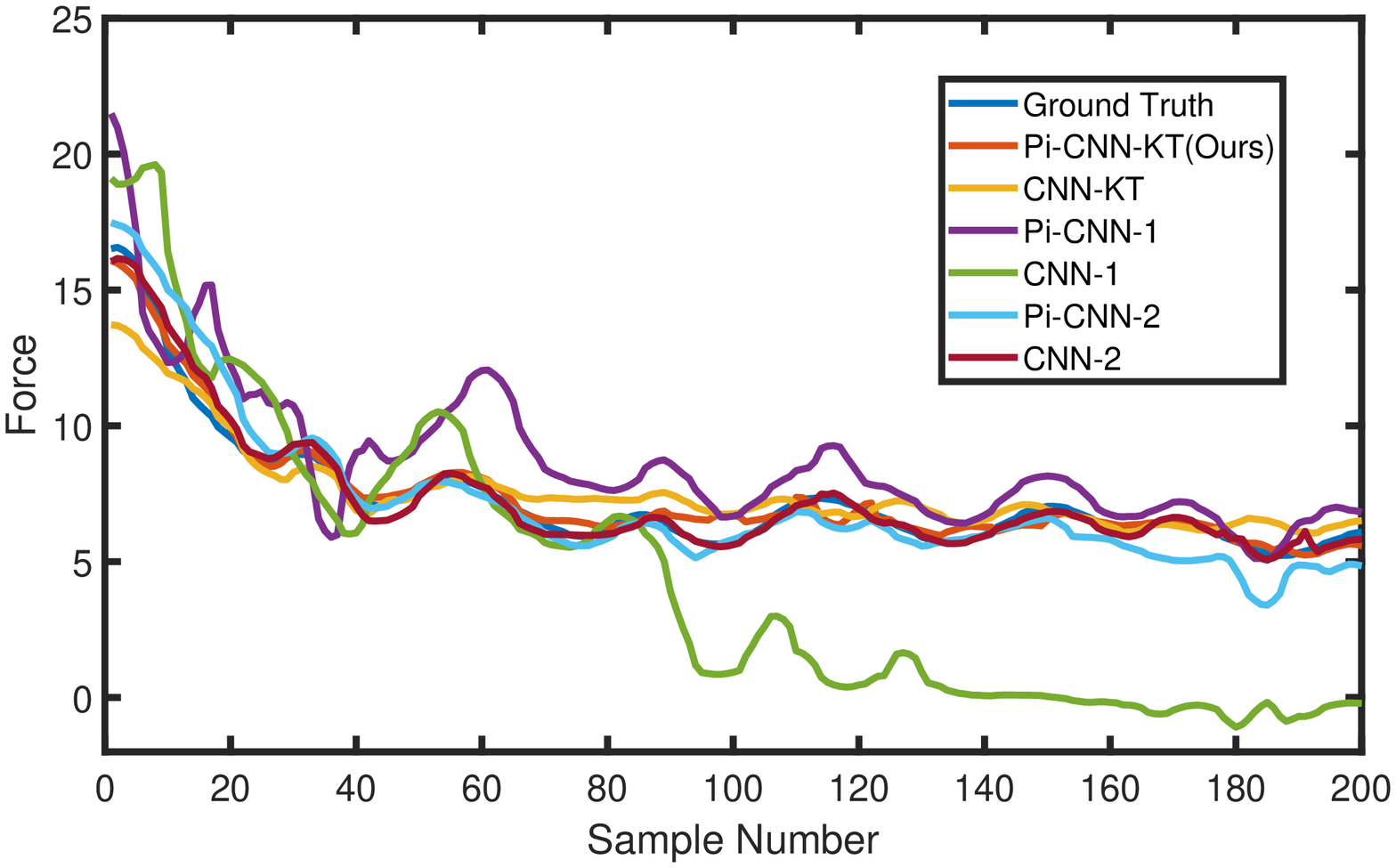}
\centerline{{\fontsize{7.5pt}{9.8pt}\selectfont (f) Predicted results of ECU}}
\end{minipage}\\
\caption{Representative predicted results of the proposed framework and baseline methods in the multiple-to-single scenario. The personalised network is trained by the data from one specific subject, while the rest seven subjects' data are utilised for training the generic network.}
\label{fig:20}
\end{figure*}

Fig.~\ref{fig:20} illustrates the representative predicted results of the proposed framework and baseline methods in the multiple-to-single scenario. Similar to the single-to-single scenario, comparing with baseline methods, the proposed framework could achieve more satisfactory predicted results, and fit the ground truth curves well. Table~\ref{tab:2} details subject's RMSEs and average $CC$s of the proposed framework and baseline methods, where the six predicted outputs are normalized. Table~\ref{tab:3} details output's RMSEs and average $CC$s of the proposed framework and baseline methods, in which average values of the eight subjects' predicted outputs are calculated. According to Table~\ref{tab:2} and Table~\ref{tab:3}, the proposed framework still could achieve smaller RMSEs and higher $CC$s under different subjects' data for training the generic network and the personalised network, indicating its great tracking capability. 

\begin{table}[htbp]
    {
    \caption{Subject's normalized RMSEs and average $CC$s of the proposed framework and baseline methods in the multiple-to-single scenario}
    \label{tab:2}
    \centering
    \resizebox{\linewidth}{!}
    {
    \begin{tabular}{c|c|cccccccc}
    \toprule
    \bfseries Metrics & \bfseries Methods  & \bfseries S1 & \bfseries S2 & \bfseries S3 & \bfseries S4 & \bfseries S5 & \bfseries S6 & \bfseries S7 & \bfseries S8\\
    \midrule
    \multirow{6}{*}{\bfseries } &  \textbf{Pi-CNN-KT} & \textbf{0.10} & \textbf{0.10} & \textbf{0.09} & \textbf{0.10} & \textbf{0.08} & \textbf{0.10} & \textbf{0.11} & \textbf{0.09}\\
    \cmidrule{2-10}
   \bfseries Normalized & CNN-KT & 0.13 & 0.11 & 0.09 & 0.13 & 0.12 & 0.12 & 0.14 & 0.12\\
    \cmidrule{2-10} 
 \bfseries RMSE & Pi-CNN-1 & 0.52 & 0.55 & 0.50 & 0.54 & 0.54 & 0.51 & 0.52 & 0.51\\
    \cmidrule{2-10} 
   & CNN-1 & 0.43 & 0.41 & 0.39 & 0.42 & 0.43 & 0.46 & 0.45 & 0.45\\
    \cmidrule{2-10}
    & Pi-CNN-2 & 0.23 & 0.24 & 0.20 & 0.27 & 0.22 & 0.22 & 0.20 & 0.19\\
    \cmidrule{2-10} 
    & CNN-2 & 0.19 & 0.17 & 0.20 & 0.15 & 0.17 & 0.16 & 0.17 & 0.16\\
   \midrule
    \multirow{6}{*}{\bfseries $CC$} &  \textbf{Pi-CNN-KT} & \textbf{0.97} & \textbf{0.96} & \textbf{0.97} & \textbf{0.98} & \textbf{0.98} & \textbf{0.97} & \textbf{0.96} & \textbf{0.98}\\
    \cmidrule{2-10} 
    & CNN-KT & 0.95 & 0.96 & 0.95 & 0.94 & 0.97 & 0.94  & 0.93 & 0.96\\
    \cmidrule{2-10} 
    & Pi-CNN-1 & 0.55 & 0.57 & 0.56 & 0.53 & 0.54 & 0.55 & 0.53 & 0.53\\
    \cmidrule{2-10} 
    & CNN-1 & 0.57 & 0.55 & 0.51 & 0.55 & 0.53 & 0.56 & 0.57 & 0.54\\
    \cmidrule{2-10} 
    & Pi-CNN-2 & 0.90 & 0.93 & 0.91 & 0.91 & 0.90 & 0.92 & 0.92 & 0.92\\
    \cmidrule{2-10} 
    & CNN-2 & 0.91 & 0.91 & 0.90 & 0.92 & 0.91 & 0.88 & 0.89 & 0.93\\
      \bottomrule
    \end{tabular}}}
\end{table}

\begin{table}[htbp]
    {
    \caption{Output's average RMSEs and average $CC$s of the proposed framework and baseline methods in the multiple-to-single scenario}
    \label{tab:3}
    \centering
    \resizebox{\linewidth}{!}
    {
    \begin{tabular}{c|c|cccccc}
    \toprule
    \bfseries Metrics & \bfseries Methods & \bfseries Angle  & \bfseries FCR & \bfseries FCU & \bfseries ECRL & \bfseries ECRB & \bfseries ECU \\
    \midrule
    \multirow{6}{*}{\bfseries } & \textbf{Pi-CNN-KT} & \textbf{10.27} & \textbf{5.19} & \textbf{4.61} & \textbf{1.12} & \textbf{1.27} & \textbf{0.42} \\
    \cmidrule{2-8} 
   & CNN-KT & 14.40 & 5.71 & 6.17 & 1.51 & 1.35 & 0.71 \\
    \cmidrule{2-8} 
  \bfseries Average &  Pi-CNN-1 & 41.21 & 15.41 & 31.33 & 23.03 & 11.12 & 2.76\\
    \cmidrule{2-8} 
    \bfseries RMSE  & CNN-1 & 38.39 & 15.59 & 29.02 & 8.87 & 13.45 & 5.16 \\
    \cmidrule{2-8}
    & Pi-CNN-2 & 26.41 & 7.38 & 8.41 & 1.75 & 2.18 & 1.02 \\
    \cmidrule{2-8} 
    & CNN-2 & 20.23 & 7.49 & 7.55 & 1.53 & 1.63 & 1.49 \\
   \midrule
    \multirow{6}{*}{\bfseries $CC$} &  \textbf{Pi-CNN-KT} & \textbf{0.95} & \textbf{0.94} & \textbf{0.97} & \textbf{0.99} & \textbf{0.98} & \textbf{0.99} \\
    \cmidrule{2-8} 
    & CNN-KT & 0.93 & 0.92 & 0.95 & 0.99 & 0.97 & 0.98 \\
    \cmidrule{2-8} 
    & Pi-CNN-1 & 0.78 & 0.14 & -0.17 & 0.85 & 0.83 & 0.76 \\
    \cmidrule{2-8} 
    & CNN-1 & 0.80 & 0.13 & 0.51 & 0.87 & 0.05 & 0.85 \\
    \cmidrule{2-8} 
    & Pi-CNN-2 & 0.86 & 0.86 & 0.92 & 0.99 & 0.96 & 0.97 \\
    \cmidrule{2-8} 
    & CNN-2 & 0.89 & 0.86 & 0.94 & 0.99 & 0.95 & 0.95 \\
      \bottomrule
    \end{tabular}}}
\end{table}

\section{Discussions}
\label{sec:discussion}
In this section, effects of dataset sizes and time consumption in physics-informed knowledge transfer phase are first analyzed, and then essential advantages of physics-informed deep transfer learning in facilitating the musculoskeletal modelling personalisation and flexibility of the proposed physics-informed deep learning framework are discussed.

\subsection{Effects of Dataset Sizes on Physics-informed Knowledge Transfer} 
Table~\ref{tab:4} depicts the RMSEs of the proposed framework under different training dataset sizes in the knowledge transfer phase, from 20\% to 80\% of the data from the new subject, used for fine-tuning the personalised network both in the single-to-single and the multiple-to-single scenarios. Observed from Table~\ref{tab:4}, the predicted performance of the proposed framework is relatively stable with the increase of the number of the data used for knowledge transfer, which indicates that the proposed framework is not sensitive to the training dataset sizes during the knowledge transfer phase.

\begin{table}[htbp]
    \caption{RMSEs of the proposed framework under different dataset sizes in the knowledge transfer phase}
    \label{tab:4}
    \centering
    \resizebox{0.8\linewidth}{!}
    {
    \begin{tabular}{c|c|cccc}
    \toprule
    \bfseries Scenarios & \bfseries Outputs & \bfseries 20\%  & \bfseries 40\% & \bfseries 60\% & \bfseries 80\% \\
    \midrule
    \multirow{6}{*}{\bfseries Single } & Angle & 10.26 & 11.18 &  10.67 & 10.59 \\
    \cmidrule{2-6} 
     & FCR & 4.75 & 4.51 & 4.45  & 4.63  \\
    \cmidrule{2-6} 
   &  FCU & 3.22 & 3.03 & 2.82  & 3.67  \\
    \cmidrule{2-6} 
   & ECRL & 2.16 & 3.27 & 3.13  & 4.11 \\
    \cmidrule{2-6} 
    & ECRB & 2.56 & 2.58 & 2.45  & 2.15  \\
    \cmidrule{2-6} 
    & ECU & 0.72 & 0.59 & 0.61  & 0.79  \\
   \midrule
    \multirow{6}{*}{\bfseries Multiple} & Angle & 10.25 & 9.52 & 8.81 & 9.22\\
    \cmidrule{2-6} 
    & FCR & 5.71 & 5.23 & 4.97 & 4.21 \\
    \cmidrule{2-6} 
    &  FCU & 6.17 & 5.51 & 5.43 & 5.26  \\
    \cmidrule{2-6} 
    & ECRL & 0.98 & 0.92 & 1.03 & 1.05  \\
    \cmidrule{2-6} 
    & ECRB & 1.27 & 1.23 & 1.25 & 1.36  \\
    \cmidrule{2-6} 
    & ECU & 0.61 & 0.47 & 0.60 & 0.63  \\
      \bottomrule
    \end{tabular}}
\end{table}

\subsection{Time Consumption of Physics-informed Knowledge Transfer} 
Table~\ref{tab:5} lists the time consumption of the proposed framework, Pi-CNN-1, CNN-1, and CNN-KT both in the single-to-single and the multiple-to-single scenarios. Accordingly, the time consumption of the proposed framework is much less than Pi-CNN-1 and CNN-1, especially in the multiple-to-single scenario. Because both Pi-CNN-1 and CNN-1 need to train the data-driven model using the whole data, while the proposed framework only utilises part of the data from the new subject to optimise the parameters relating to the subject-specific information in the personalised network. Additionally, the time consumption of CNN-KT is less than the proposed framework, it because that the physics-based domain knowledge is embedded into the loss function to penalise/regularise the neural network during the knowledge transfer phase, but CNN-KT only needs to minimise the MSE loss. 
 
\begin{table}[htbp]
    {
    \caption{Comparisons of Time Consumption (Minute)}
    \label{tab:5}
    \centering
    \resizebox{0.88\linewidth}{!}
    {
    \begin{tabular}{c|cccc}
    \toprule
    \bfseries Scenarios & \bfseries Pi-CNN-1 & \bfseries CNN-1 & \bfseries Pi-CNN-KT & \bfseries CNN-KT\\
    \midrule
    \bfseries Single & 80 & 64 & \textbf{33} & 27 \\
    \midrule
    \bfseries Multiple & 361 & 305 & \textbf{37} & 29 \\
         \bottomrule
    \end{tabular}}}
\end{table} 
   
\subsection{Physics-informed Deep Transfer Learning to Facilitate Musculoskeletal Modelling Personalisation} 
As mentioned above, the generic musculoskeletal model without particular personalisation may be enough when we want to investigate the musculoskeletal phenomena decoupled from individuals or groups, such as how muscles are neurally recruited, how muscles transfer force around multiple joints, or motor control principles, etc. However, when we study the musculoskeletal function of the specific subject, musculoskeletal model with the unique anatomy and neurophysiology of the subject is necessary~\cite{hiasa2019automated,sartori2017subject}. Additionally, understanding the mechanism underlying the specific individual's musculoskeletal function is important for rehabilitation, such as design of the personalised assistive device and human-machine interface, and formulating personalised rehabilitation intervention strategy for the specific subject based on his/her anatomy and impairment~\cite{valente2014subject,arones2020musculoskeletal,pizzolato2020targeted}. 

One of the challenging issues for musculoskeletal modelling personalisation is the shifts of statistical characteristics of the collected EMG signals from different individuals mainly caused by the diversity of the anatomical, physiological and biochemical characteristics between individuals, which may seriously degrade the performance of the conventional data-driven model~\cite{kim2019subject,chen2020hand}. Therefore, conventional data-driven methods usually require a large number of subject-specific data to retrain the existing model, which is time-consuming and labor-intensive~\cite{goswami2020transfer,chakraborty2021transfer}. In addition, data-driven methods highly rely on the quality of the collected EMG signals, and the ill-conditioned and noisy training data may impose negative and unpredictable results. Although some existing transfer learning frameworks, such as MetaSleepLearner~\cite{banluesombatkul2020metasleeplearner}, EEGWaveNet~\cite{thuwajit2021eegwavenet}, and MIN2Net~\cite{autthasan2021min2net}, could achieve satisfactory performance in subject-independent scenarios, they were developed without considering the physics-based domain knowledge, and such kind of “black-box” solutions cannot reflect the underlying physical mechanisms during the modelling process. Differently, the proposed framework is more robust and with better generalization by integrating physics-based domain knowledge that from kinematic measurements and the physical understanding of the neuromusculoskeletal coupling into the deep neural networks to reflect physical or physiological mechanisms~\cite{reinbold2021robust,wang2022and,rudin2022interpretable,yang2019adversarial}. Furthermore, less subject-specific data are required in the knowledge transfer phase due to the twin neural networks mechanism.

\subsection{Flexibility of Physics-informed Deep Learning Framework} 
The aim of our work is to develop the next-generation physics-informed data-driven musculoskeletal models, which could seamlessly integrate the existing physics-based domain knowledge into the deep learning-based data-driven models. The proposed method is only an attempt to bring physics information into data-driven models to overcome its limitations by creating data reflecting the underlying physical mechanisms, and also bring the powerful deep learning techniques into physics-based musculoskeletal models to reduce computational demands in data processing and improve execution speed for real-time applications. Moreover, the embedded physics law illustrates the relationship between sEMG, and muscle forces and joint kinematics. We choose the wrist joint as the exemplar, by considering that wrist primary muscles are superficial muscle and can be easily measured, and wrist muscle forces and joint kinematics estimation is a kind of application widely considered in the existing works. It also could be generalized to other joints and more scenarios. In this paper, we use CNN to implement the proposed physics-informed deep learning framework. The proposed framework is actually flexible, the deep neural network can be replaced depending on the specific requirements, such as LSTM and ResNet, etc.

{Table~\ref{tab:6} lists the detailed comparison results between Pi-CNN (CNN as the deep neural network in the proposed physics-informed deep learning framework) and Pi-LSTM (LSTM as the deep neural network in the proposed physics-informed deep learning framework). According to Table~\ref{tab:6}, we can find that Pi-CNN and Pi-LSTM achieve comparable performance, indicating that different types of deep neural networks can be embedded into the proposed physics-informed deep learning framework.}

\begin{table}[htbp]
    {
    \caption{{Comparisons between Pi-CNN and Pi-LSTM}}
    \label{tab:6}
    \centering
    \resizebox{0.88\linewidth}{!}
    {
    \begin{tabular}{c|cccccc}
    \toprule
    \bfseries  & \bfseries Angle & \bfseries FCR & \bfseries FCU & \bfseries ECRL & \bfseries ECRB & \bfseries ECU \\
    \midrule
    \bfseries Pi-CNN & 9.35 & 5.17 & 3.98 & 2.77 & 2.61 & 0.81 \\
    \midrule
    \bfseries Pi-LSTM & 9.92 & 5.26 & 3.76 & 2.99 & 2.52 & 0.75\\
         \bottomrule
    \end{tabular}}}
\end{table}

\section{Conclusion}
\label{sec:conclusion}
In this paper, a physics-informed deep transfer learning framework is designed for musculoskeletal modelling personalisation. Different from conventional transfer learning methods, physics-based domain knowledge is integrated into deep neural networks as soft constraints to penalise/regularise the loss function of CNNs utilised in the proposed framework. The embedded physics-based domain knowledge could help strengthen robustness and generalization, and the number of required subject-specific data is reduced during the knowledge transfer phase. Moreover, the utility of the twin neural networks mechanism significantly reduces computational burdens in personalised network fine-tuning. Comprehensive experiments on wrist muscle forces and joint angle estimation demonstrate the effectiveness of the proposed framework. It should be noted that, in this study, we assume all the data used in the experiments are labelled, but the available data of the new subject may be unlabelled in clinical applications. Therefore, we will focus on the future physics-informed unsupervised transfer learning and physics-informed semi-supervised transfer learning strategies.





\bibliographystyle{IEEEtran}
\bibliography{IEEEabrv,reference}

\begin{thebibliography}{10}
\providecommand{\url}[1]{#1}
\csname url@samestyle\endcsname
\providecommand{\newblock}{\relax}
\providecommand{\bibinfo}[2]{#2}
\providecommand{\BIBentrySTDinterwordspacing}{\spaceskip=0pt\relax}
\providecommand{\BIBentryALTinterwordstretchfactor}{4}
\providecommand{\BIBentryALTinterwordspacing}{\spaceskip=\fontdimen2\font plus
\BIBentryALTinterwordstretchfactor\fontdimen3\font minus
  \fontdimen4\font\relax}
\providecommand{\BIBforeignlanguage}[2]{{%
\expandafter\ifx\csname l@#1\endcsname\relax
\typeout{** WARNING: IEEEtran.bst: No hyphenation pattern has been}%
\typeout{** loaded for the language `#1'. Using the pattern for}%
\typeout{** the default language instead.}%
\else
\language=\csname l@#1\endcsname
\fi
#2}}
\providecommand{\BIBdecl}{\relax}
\BIBdecl

\bibitem{li2022new}
Q.~Li, Z.~Luo, and J.~Zheng, ``A new deep anomaly detection-based method for
  user authentication using multichannel surface {EMG} signals of hand
  gestures,'' \emph{IEEE Trans. Instrum. Meas.}, vol.~71, pp. 1--11, 2022.

\bibitem{bennett2022emg}
K.~J. Bennett, C.~Pizzolato, S.~Martelli, J.~S. Bahl, A.~Sivakumar, G.~J.
  Atkins, L.~B. Solomon, and D.~Thewlis, ``{EMG}-informed neuromusculoskeletal
  models accurately predict knee loading measured using instrumented
  implants,'' \emph{IEEE Trans. Biomed. Eng.}, 2022.

\bibitem{weng2022adaptive}
J.~Weng, E.~Hashemi, and A.~Arami, ``Adaptive reference inverse optimal control
  for natural walking with musculoskeletal models,'' \emph{IEEE Trans. Neural
  Syst. Rehabil. Eng.}, 2022.

\bibitem{jung2021intramuscular}
M.~K. Jung, S.~Muceli, C.~Rodrigues, {\'A}.~Meg{\'\i}a-Garc{\'\i}a,
  A.~Pascual-Valdunciel, A.~J. Del-Ama, A.~Gil-Agudo, J.~C. Moreno, F.~O.
  Barroso, J.~L. Pons, and D.~Farina, ``Intramuscular {EMG}-driven
  musculoskeletal modelling: Towards implanted muscle interfacing in spinal
  cord injury patients,'' \emph{{IEEE} Trans. Biomed. Eng.}, vol.~69, no.~1,
  pp. 63--74, 2021.

\bibitem{zhao2020emg}
Y.~Zhao, Z.~Zhang, Z.~Li, Z.~Yang, A.~A. Dehghani-Sanij, and S.~Xie, ``An
  {EMG}-driven musculoskeletal model for estimating continuous wrist motion,''
  \emph{{IEEE} Trans. Neural Syst. Rehabil. Eng.}, vol.~28, no.~12, pp.
  3113--3120, 2020.

\bibitem{xiong2021deep}
D.~Xiong, D.~Zhang, X.~Zhao, and Y.~Zhao, ``Deep learning for {EMG}-based
  human-machine interaction: A review,'' \emph{IEEE/CAA J. Autom. Sin.},
  vol.~8, no.~3, pp. 512--533, 2021.

\bibitem{rane2019deep}
L.~Rane, Z.~Ding, A.~H. McGregor, and A.~M. Bull, ``Deep learning for
  musculoskeletal force prediction,'' \emph{Annu. Rev. Biomed. Eng.}, vol.~47,
  no.~3, pp. 778--789, 2019.

\bibitem{de2020machine}
A.~De~Brabandere, J.~Emmerzaal, A.~Timmermans, I.~Jonkers, B.~Vanwanseele, and
  J.~Davis, ``A machine learning approach to estimate hip and knee joint
  loading using a mobile phone-embedded {IMU},'' \emph{Front. Bioeng.
  Biotechnol.}, vol.~8, p. 320, 2020.

\bibitem{makaram2021analysis}
N.~Makaram, P.~A. Karthick, and R.~Swaminathan, ``Analysis of dynamics of {EMG}
  signal variations in fatiguing contractions of muscles using transition
  network approach,'' \emph{IEEE Trans. Instrum. Meas.}, vol.~70, pp. 1--8,
  2021.

\bibitem{zhang2020robust}
J.~Zhang, Y.~Li, W.~Xiao, and Z.~Zhang, ``Robust extreme learning machine for
  modeling with unknown noise,'' \emph{J. Franklin Inst.}, vol. 357, no.~14,
  pp. 9885--9908, 2020.

\bibitem{bao2020cnn}
T.~Bao, S.~A.~R. Zaidi, S.~Xie, P.~Yang, and Z.-Q. Zhang, ``A {CNN}-{LSTM}
  hybrid model for wrist kinematics estimation using surface
  electromyography,'' \emph{IEEE Trans. Instrum. Meas.}, vol.~70, pp. 1--9,
  2020.

\bibitem{bao2022towards}
T.~Bao, S.~Q. Xie, P.~Yang, P.~Zhou, and Z.~Zhang, ``Towards robust, adaptive
  and reliable upper-limb motion estimation using machine learning and deep
  learning-{A} survey in myoelectric control,'' \emph{IEEE J. Biomed. Health
  Inform.}, 2022.

\bibitem{tang2021decoding}
X.~Tang, X.~Zhang, M.~Chen, X.~Chen, and X.~Chen, ``Decoding muscle force from
  motor unit firings using encoder-decoder networks,'' \emph{IEEE Trans. Neural
  Syst. Rehabil. Eng.}, vol.~29, pp. 2484--2495, 2021.

\bibitem{wimalasena2022estimating}
L.~N. Wimalasena, J.~F. Braun, M.~R. Keshtkaran, D.~Hofmann, J.~{\'A}. Gallego,
  C.~Alessandro, M.~C. Tresch, L.~E. Miller, and C.~Pandarinath, ``Estimating
  muscle activation from {EMG} using deep learning-based dynamical systems
  models,'' \emph{J. Neural Eng.}, vol.~19, no.~3, p. 036013, 2022.

\bibitem{burton2021machine}
W.~S. Burton~II, C.~A. Myers, and P.~J. Rullkoetter, ``Machine learning for
  rapid estimation of lower extremity muscle and joint loading during
  activities of daily living,'' \emph{J. Biomech.}, vol. 123, p. 110439, 2021.

\bibitem{dao2019deep}
T.~T. Dao, ``From deep learning to transfer learning for the prediction of
  skeletal muscle forces,'' \emph{Med. Biol. Eng. Comput.}, vol.~57, no.~5, pp.
  1049--1058, 2019.

\bibitem{bao2021inter}
T.~Bao, S.~A.~R. Zaidi, S.~Xie, P.~Yang, and Z.-Q. Zhang, ``Inter-subject
  domain adaptation for {CNN}-based wrist kinematics estimation using s{EMG},''
  \emph{IEEE Trans. Neural Syst. Rehabil. Eng.}, vol.~29, pp. 1068--1078, 2021.

\bibitem{wang2018sensor}
W.~Wang, B.~Chen, P.~Xia, J.~Hu, and Y.~Peng, ``Sensor fusion for myoelectric
  control based on deep learning with recurrent convolutional neural
  networks,'' \emph{Artif. Organs}, vol.~42, no.~9, pp. E272--E282, 2018.

\bibitem{kim2019subject}
K.-T. Kim, C.~Guan, and S.-W. Lee, ``A subject-transfer framework based on
  single-trial {EMG} analysis using convolutional neural networks,'' \emph{IEEE
  Trans. Neural Syst. Rehabil. Eng.}, vol.~28, no.~1, pp. 94--103, 2019.

\bibitem{sartori2016neural}
M.~Sartori, D.~G. Llyod, and D.~Farina, ``Neural data-driven musculoskeletal
  modeling for personalized neurorehabilitation technologies,'' \emph{{IEEE}
  Trans. Biomed. Eng.}, vol.~63, no.~5, pp. 879--893, 2016.

\bibitem{karniadakis2021physics}
G.~E. Karniadakis, I.~G. Kevrekidis, L.~Lu, P.~Perdikaris, S.~Wang, and
  L.~Yang, ``Physics-informed machine learning,'' \emph{Nat. Rev. Phys.},
  vol.~3, no.~6, pp. 422--440, 2021.

\bibitem{raissi2019physics}
M.~Raissi, P.~Perdikaris, and G.~E. Karniadakis, ``Physics-informed neural
  networks: A deep learning framework for solving forward and inverse problems
  involving nonlinear partial differential equations,'' \emph{J. Comput.
  Phys.}, vol. 378, pp. 686--707, 2019.

\bibitem{willard2021integrating}
J.~Willard, X.~Jia, S.~Xu, M.~Steinbach, and V.~Kumar, ``Integrating scientific
  knowledge with machine learning for engineering and environmental systems,''
  \emph{ACM Comput. Surv.}, 2021.

\bibitem{zhao2022adaptive}
Y.~Zhao, K.~Qian, S.~Bo, Z.~Zhang, Z.~Li, G.-Q. Li, A.~A. Dehghani-Sanij, and
  S.~Q. Xie, ``Adaptive cooperative control strategy for a wrist exoskeleton
  using model-based joint impedance estimation,'' \emph{IEEE/ASME Trans.
  Mechatron.}, 2022.

\bibitem{zhang2020non}
J.~Zhang, Y.~Li, W.~Xiao, and Z.~Zhang, ``Non-iterative and fast deep learning:
  Multilayer extreme learning machines,'' \emph{J. Franklin Inst.}, vol. 357,
  no.~13, pp. 8925--8955, 2020.

\bibitem{hiasa2019automated}
Y.~Hiasa, Y.~Otake, M.~Takao, T.~Ogawa, N.~Sugano, and Y.~Sato, ``Automated
  muscle segmentation from clinical {CT} using {B}ayesian {U}-{N}et for
  personalized musculoskeletal modeling,'' \emph{IEEE Trans. Med. Imaging},
  vol.~39, no.~4, pp. 1030--1040, 2019.

\bibitem{sartori2017subject}
M.~Sartori, J.~Rubenson, D.~G. Lloyd, D.~Farina, and F.~A. Panizzolo,
  ``Subject-specificity via 3{D} ultrasound and personalized musculoskeletal
  modeling,'' in \emph{Converging Clinical and Engineering Research on
  Neurorehabilitation II}.\hskip 1em plus 0.5em minus 0.4em\relax Springer,
  2017, pp. 639--642.

\bibitem{valente2014subject}
G.~Valente, L.~Pitto, D.~Testi, A.~Seth, S.~L. Delp, R.~Stagni, M.~Viceconti,
  and F.~Taddei, ``Are subject-specific musculoskeletal models robust to the
  uncertainties in parameter identification?'' \emph{PLoS One}, vol.~9, no.~11,
  p. e112625, 2014.

\bibitem{arones2020musculoskeletal}
M.~M. Arones, M.~S. Shourijeh, C.~Patten, and B.~J. Fregly, ``Musculoskeletal
  model personalization affects metabolic cost estimates for walking,''
  \emph{Front. Bioeng. Biotechnol.}, vol.~8, p. 588925, 2020.

\bibitem{pizzolato2020targeted}
C.~Pizzolato, V.~B. Shim, D.~G. Lloyd, D.~Devaprakash, S.~J. Obst,
  R.~Newsham-West, D.~F. Graham, T.~F. Besier, M.~H. Zheng, and R.~S. Barrett,
  ``Targeted achilles tendon training and rehabilitation using personalized and
  real-time multiscale models of the neuromusculoskeletal system,''
  \emph{Front. Bioeng. Biotechnol.}, p. 878, 2020.

\bibitem{chen2020hand}
X.~Chen, Y.~Li, R.~Hu, X.~Zhang, and X.~Chen, ``Hand gesture recognition based
  on surface electromyography using convolutional neural network with transfer
  learning method,'' \emph{IEEE J. Biomed. Health Inform.}, vol.~25, no.~4, pp.
  1292--1304, 2020.

\bibitem{goswami2020transfer}
S.~Goswami, C.~Anitescu, S.~Chakraborty, and T.~Rabczuk, ``Transfer learning
  enhanced physics informed neural network for phase-field modeling of
  fracture,'' \emph{Theor. Appl. Fract. Mech.}, vol. 106, p. 102447, 2020.

\bibitem{chakraborty2021transfer}
S.~Chakraborty, ``Transfer learning based multi-fidelity physics informed deep
  neural network,'' \emph{J. Comput. Phys.}, vol. 426, p. 109942, 2021.

\bibitem{banluesombatkul2020metasleeplearner}
N.~Banluesombatkul, P.~Ouppaphan, P.~Leelaarporn, P.~Lakhan, B.~Chaitusaney,
  N.~Jaimchariyatam, E.~Chuangsuwanich, W.~Chen, H.~Phan, N.~Dilokthanakul
  \emph{et~al.}, ``Meta{S}leep{L}earner: A pilot study on fast adaptation of
  bio-signals-based sleep stage classifier to new individual subject using
  meta-learning,'' \emph{IEEE J. Biomed. Health Inform.}, vol.~25, no.~6, pp.
  1949--1963, 2020\color{black}.

\bibitem{thuwajit2021eegwavenet}
P.~Thuwajit, P.~Rangpong, P.~Sawangjai, P.~Autthasan, R.~Chaisaen,
  N.~Banluesombatkul, P.~Boonchit, N.~Tatsaringkansakul, T.~Sudhawiyangkul, and
  T.~Wilaiprasitporn, ``{EEGW}ave{N}et: Multiscale {CNN}-based spatiotemporal
  feature extraction for {EEG} seizure detection,'' \emph{IEEE Trans. Ind.
  Inform.}, vol.~18, no.~8, pp. 5547--5557, 2021\color{black}.

\bibitem{autthasan2021min2net}
P.~Autthasan, R.~Chaisaen, T.~Sudhawiyangkul, P.~Rangpong, S.~Kiatthaveephong,
  N.~Dilokthanakul, G.~Bhakdisongkhram, H.~Phan, C.~Guan, and
  T.~Wilaiprasitporn, ``{MIN}2{N}et: End-to-end multi-task learning for
  subject-independent motor imagery {EEG} classification,'' \emph{IEEE Trans.
  Biomed. Eng.}, vol.~69, no.~6, pp. 2105--2118, 2021\color{black}.

\bibitem{reinbold2021robust}
P.~A. Reinbold, L.~M. Kageorge, M.~F. Schatz, and R.~O. Grigoriev, ``Robust
  learning from noisy, incomplete, high-dimensional experimental data via
  physically constrained symbolic regression,'' \emph{Nat. Commun.}, vol.~12,
  no.~1, pp. 1--8, 2021.

\bibitem{wang2022and}
S.~Wang, X.~Yu, and P.~Perdikaris, ``When and why {PINN}s fail to train: A
  neural tangent kernel perspective,'' \emph{J. Comput. Phys.}, vol. 449, pp.
  1--28, 2022.

\bibitem{rudin2022interpretable}
C.~Rudin, C.~Chen, Z.~Chen, H.~Huang, L.~Semenova, and C.~Zhong,
  ``Interpretable machine learning: Fundamental principles and 10 grand
  challenges,'' \emph{Stat. Surv.}, vol.~16, pp. 1--85, 2022.

\bibitem{yang2019adversarial}
Y.~Yang and P.~Perdikaris, ``Adversarial uncertainty quantification in
  physics-informed neural networks,'' \emph{J. Comput. Phys.}, vol. 394, pp.
  136--152, 2019.

\end{thebibliography}


%

%





\end{document}